\documentclass[aps,prl,floatfix,twocolumn,showpacs,10pt,longbibliography]{revtex4-2}
\usepackage{amsmath}
\usepackage{braket} 
\usepackage{url}
\usepackage[ruled,vlined]{algorithm2e}
\usepackage{algorithmic}
\usepackage{graphicx}
\usepackage{dcolumn}
\usepackage{bm}
\usepackage{amssymb}
\usepackage{rotating}
\usepackage[abs]{overpic}
\usepackage{xcolor}
\usepackage{booktabs}
\usepackage{tikz}
\usepackage{tabularx}
\usepackage{ragged2e}
\usepackage{hyperref}
\hypersetup{colorlinks = true, citebordercolor={blue}, linkcolor={blue}, citecolor={blue}, urlcolor={blue}}

\begin{document}


\title{Fewer measurements from shadow tomography with $N$-representability conditions}
\author{Irma Avdic and David A. Mazziotti}

\email{damazz@uchicago.edu}

\affiliation{Department of Chemistry and The James Franck Institute, The University of Chicago, Chicago, IL 60637 USA}

\date{Submitted December 18, 2023}

\begin{abstract}
Classical shadow tomography provides a randomized scheme for approximating the quantum state and its properties at reduced computational cost with applications in quantum computing.  In this Letter we present an algorithm for realizing fewer measurements in the shadow tomography of many-body systems by imposing $N$-representability constraints.  Accelerated tomography of the two-body reduced density matrix (2-RDM) is achieved by combining classical shadows with necessary constraints for the 2-RDM to represent an $N$-body system, known as $N$-representability conditions.  We compute the ground-state energies and 2-RDMs of hydrogen chains and the N$_{2}$ dissociation curve.  Results demonstrate a significant reduction in the number of measurements with important applications to quantum many-body simulations on near-term quantum devices. 
\end{abstract}

\maketitle


\textit{Introduction.---} Shadow tomography~\cite{Aaronson2020, Huang2020, Huang2022, Hu2023, McGinley2022, OGorman2022, Low2022} has recently emerged as a critical technique in quantum information theory, offering an efficient implementation of random sampling for characterizing complicated quantum systems.  Built around the principle of using minimal measurements to image a quantum state~\cite{Zhao2021, Peng2023}, this approach stands at the forefront of recent advances in tomography~\cite{Lanyon2017, Yen2020, Reagor2018, Cramer2010, Rambach2021, ODonnell2016, Haah2017,Gupta2021, Gupta2022, Gard2020, Tang2021} that aim to address outstanding challenges in quantum computing~\cite{Corcoles.2020} and quantum state verification~\cite{Jiang.2020}.  Despite its promise and generality, however, shadow tomography has limitations, including its scalability for measuring large quantum states and its robustness with noisy data as generated by near-term intermediate-scale quantum (NISQ) devices.  These limitations can be particularly challenging for many-body quantum systems whose complexity increases dramatically with system size~\cite{mcardle_2020, Head-Marsden2020, Smart2021_2, Benavides-Riveros.2022}. \\


In this Letter, we present a substantial acceleration of the shadow tomography of two-particle reduced density matrices (2-RDM) of many-body systems by imposing $N$-representability constraints~\cite{M2007, Coleman2000, Coleman1963, Garrod.1964, Kummer1967, Erdahl1978, mazziotti2001,  Mazziotti2006_2, Mazziotti2012, Mazziotti2023}.  Conventional shadow tomography neglects the fact that a 2-RDM must obey significant, non-trivial constraints, known as $N$-representability conditions~\cite{M2007, Coleman2000}, to ensure that it derives from the integration of at least one valid $N$-particle density matrix.  Here, we introduce a classical algorithm that reconstructs the 2-RDM from a set of classical shadows while enforcing necessary $N$-representability conditions.  In particular, we consider a hierarchy of $N$-representability conditions on the 2-RDM that arise from constraining $(p+1)$ $p$-particle metric matrices to be positive semidefinite, known as the $p$-positivity conditions~\cite{mazziotti2001, Mazziotti2006_2, Mazziotti2012, Mazziotti2023}.  Both the positivity conditions and the shadow constraints can be imposed on the 2-RDM through a special family of convex optimization, known as semidefinite programming~\cite{VB1996, Mazziotti2004, M2011}.  We can also view the approach as adding the shadow constraints obtained from measuring a wave function (e.g., a wave function prepared on a quantum computer) to the variational calculation of the 2-RDM subject to $N$-representability conditions~\cite{Nakata.2001, Mazziotti.2002kyg, Mazziotti2004, Zhao.2004, Cances.2006, Gidofalvi.2008, Shenvi.2010, Verstichel.2011, Baumgratz.2012, Mazziotti.2016co, Alcoba.2018, Mazziotti.20206vx, Li.2021, Piris.2021, Knight.2022}---a classical-computing method that has been successfully applied to strongly correlated many-electron systems from spin systems to molecules and materials~\cite{Schlimgen.2016, Boyn.2020, Kawamura.2020jms, Xie.20228s, Schouten.2023}. \\

After we develop the theory for $N$-representability-enhanced shadow tomography, we demonstrate its advantages by computing the ground-state energies and 2-RDMs of hydrogen chains and the N$_{2}$ dissociation curve.  Results reveal a significant reduction in the number of measurements for 2-RDM tomography, even in the presence of noise, with applications to many-body simulations on noisy intermediate-scale quantum devices.  \\

\textit{Theory.---}
We consider an $N$-electron system whose 2-RDM elements arise from the integration of the $N$-particle density matrix over all particles except two and can be expressed as
\begin{equation}
    ^2D^{ij}_{kl} = \bra{\Psi}\hat{a}^\dagger_{i}\hat{a}^\dagger_{j}\hat{a}^{}_{l}\hat{a}^{}_{k}\ket{\Psi},
\end{equation}
where $\hat{a}_{i}^{\dagger}$ creates a particle in orbital $i$ and $\hat{a}^{}_{i}$ annihilates a particle in orbital $i$.
The protocol developed by Huang et al.~\cite{Huang2020}, previously introduced by Aaronson~\cite{Aaronson2020}, illustrates a method for creating a classical shadow of a quantum state $\ket{\Psi}$ by first applying a unitary transformation on the state such that $\ket{\Psi} \mapsto \hat{U}\ket{\Psi}$ where $\hat{U}$ is a random matrix from an ensemble $\mathcal{U}$ and then measuring the state in the computational basis.  After passing $\ket{\Psi}$ through a unitary transformation, i.e., a quantum circuit, the classical shadow description of the 2-RDM is expressible as
\begin{equation}
    S_{n}^{pq} = \bra{\Psi}\hat{U}^{\dagger}_{n}\hat{a}^\dagger_{p}\hat{a}^\dagger_{q}\hat{a}^{}_{q}\hat{a}^{}_{p}\hat{U}_n\ket{\Psi},
    \label{eq:shadow} 
\end{equation}
where $\hat{U}_{n} = \exp{({\sum_{uv}A^{uv}_{n}\hat{a}^{\dagger}_{u}\hat{a}^{}_{v}})}$ in which the indices denote spin orbitals, $A_{n}$ is a one-body anti-Hermitian matrix, and $n$ is the index for the shadow.  The $n^{\rm th}$ shadow corresponds to the measurement of all diagonal elements of the 2-RDM after application of the $n^{\rm th}$ one-body unitary transformation $\hat{U}_{n}$ to the wave function. Each one-body unitary transformation, generated by random sampling with the Haar measure~\cite{Haar1933}, has the effect of rotating the orbitals into a new basis. From Eq.~(\ref{eq:shadow}), the collection of shadows can be written in terms of the elements of the 2-RDM as
\begin{equation}
    S^{pq}_{n} = \sum_{ijkl}{U_{n}^{pi} U_{n}^{pj} \, {^2D^{ij}_{kl}} \, U_{n}^{ql} U_{n}^{qk}}, 
\end{equation}
where $U_{n} = \exp{(A_{n})}$.  Because a sufficiently large set of the above shadows defines a system of equations that determines the 2-RDM, the technique can be used to predict the 2-RDM---or any one- or two-body expectation value---of the original state. Nonetheless, since the output state is constructed from the measurements of only classical parts of the 2-RDM, which are commutable and, hence, measurable in parallel, the tomography cost is significantly reduced. \\

Here, we use necessary $N$-representability conditions in combination with the linear constraints arising from the classical shadows to determine the 2-RDM.  The $N$-representability conditions define a convex set ${^{N}_{2}\Tilde{P}}$ of approximately $N$-representable 2-RDMs that contains the set ${^{N}_{2}{P}}$ of exactly $N$-representable 2-RDMs~\cite{Mazziotti2012}.  Likewise, a collection of $m$ classical shadow constraints define a convex set $\Tilde{S}_{m}$ of 2-RDMs that approximate the 2-RDM of the quantum state.  As the number $m$ of randomly sampled classical shadows increases, the set $\Tilde{S}_{m}$ decreases until at some limiting $m^{*}$ the set contains only the state's 2-RDM.  However, to maximize the efficiency of the tomography, we want to obtain a good approximation of the 2-RDM for $m \ll m^{*}$.  Importantly, we can significantly accelerate the convergence of the classical shadows to the correct 2-RDM by considering the intersection of the $m$-shadow set of 2-RDMs $\Tilde{S}_{m}$ with the set ${^{N}_{2}\Tilde{P}}$ of approximately $N$-representable 2-RDMs.  Because the intersection set $\Tilde{S}_{m}^{+}$ is a convex subset of $\Tilde{S}_{m}$
\begin{equation}
\Tilde{S}_{m}^{+} = \Tilde{S}_{m} \cap {^{N}_{2}\Tilde{P}} \subset \Tilde{S}_{m} ,
\end{equation}
the approximations of the quantum state's 2-RDM in $\Tilde{S}_{m}^{+}$ converges much faster with $m$ towards the correct 2-RDM than the 2-RDM approximations in the set $\Tilde{S}_{m}$ using only conventional shadow tomography.  The difference in the sets and, hence, the degree of convergence acceleration is especially pronounced for our target range of $m \ll m^{*}$.  \\

To obtain a unique 2-RDM for a finite value of $m$, we score the 2-RDMs in the convex set $\Tilde{S}_{m}^{+}$ by a merit function. To obtain a convex optimization problem, we can select the merit function to be any convex function of the 2-RDM. Potential merit functions include the nuclear norm or the Frobenius norm, as in matrix completion theory~\cite{Candes.2009, Cai.2010}. Here, we choose the merit function to be the expectation value of the energy, which, for a quantum many-body system with at most pairwise interactions, is expressible as a linear functional of the 2-RDM. We have the following convex optimization problem 
\begin{align}
    &\min_{^2D \in {^{N}_{2}\Tilde{P}}} \hspace{0.2cm} F[^2D] \\
\text{such that~~} S^n_{pq} &= ((U\otimes U)^2D(U\otimes U)^T)^{pq}_{pq},
\end{align}
for $n \in [0,m]$ where the 2-RDM is approximately $N$-representable. The optimization converges to the 2-RDM, subject to both the $N$-representability conditions and the shadow constraints, that minimizes the energy. The choice of the energy for the merit function is particularly attractive because, in the absence of the shadow constraints (i.e., $m=0$), the optimization becomes a variational 2-RDM calculation in which the energy is minimized over the approximately $N$-representable set ${^{N}_{2}\Tilde{P}}$ of 2-RDMs. From this perspective, the classical shadow constraints can be viewed as state-specific conditions that improve the performance of variational 2-RDM theory. \\

In practice, the set ${^{N}_{2}\Tilde{P}}$ can be described by a systematic hierarchy of $N$-representability conditions known as the $p$-positivity conditions~\cite{mazziotti2001, Mazziotti2006_2, Mazziotti2012, Mazziotti2023}.  The $p$-positivity conditions constrain $(p+1)$ metric matrices, which interrelate by linear mappings and contract to the 2-RDM, to be positive semidefinite.  The convex optimization becomes a semidefinite program (SDP)~\cite{VB1996, Mazziotti2004, M2011}.  If we consider only the 2-positivity conditions, we obtain the following SDP   
\begin{align}
    \min_{^2D}& \hspace{0.2cm} E[^2D] \label{eq:E} \\
\text{such that} 
    \hspace{0.2cm} ^2D &\succeq 0 \\
                   ^2Q &\succeq 0 \\
                   ^2G &\succeq 0 \\
                   {\rm Tr}(^2D) &= N(N-1) \\
                   ^2Q &= f_Q(^2D) \\
                   ^2G &= f_G(^2D) \\
                   S_n^{pq} &= ((U \otimes U)^2D(U \otimes U)^T)^{pq}_{pq}\label{eq:sdp-shadow}
\end{align}
where $^{2} Q$ and $^{2} G$ are the hole-hole and particle-hole metric matrices whose elements are given by
\begin{align}
^{2} Q^{kl}_{ij} & = \langle \Psi | {\hat a}^{}_{k} {\hat a}^{}_{l} {\hat a}^{\dagger}_{j} {\hat a}^{\dagger}_{i}| \Psi \rangle  
\label{eq:Q} \\
^{2} G^{il}_{kj} & = \langle \Psi | {\hat a}^{\dagger}_{i} {\hat a}^{}_{l} {\hat a}^{\dagger}_{j} {\hat a}^{}_{k}| \Psi \rangle , \label{eq:G} 
\end{align}
$M \succeq 0$ indicate that the matrix $M$ is constrained to be positive semidefinite, ${\rm Tr}(^2D)$ denotes the trace of the 2-RDM that is set to $N(N-1)$, and the functions $f_Q$ and $f_G$ represent the linear mappings between $^{2}Q$ and $^{2} D$ and $^{2} G$ and $^{2} D$, respectively, that are obtained from rearranging the creation and annihilation operators in Eqs.~(\ref{eq:Q}) and~(\ref{eq:G}).  Physically, the D, Q, and G conditions~\cite{M2007, Coleman1963, Garrod.1964} constrain the probability distributions of two electrons, two holes, and an electron-hole pair to be nonnegative. \\

Within the SDP protocol, we can replace the equality in Eq.~(\ref{eq:sdp-shadow}) with inequalities to offer flexibility in handling statistical error in the 2-RDM, e.g., in the presence of quantum noise on quantum devices
\begin{equation}
S_{n}^{pq} - \epsilon_{n}^{pq} \leq X^{pq}_{n} \le S_{n}^{pq} + \epsilon_{n}^{pq} \label{eq:noise1}    
\end{equation}
where
\begin{equation}
X^{pq}_{n} = ((U_{n}\otimes U_{n}) \, {{}^2D} \, ( U_{n} \otimes U_{n})^T)^{pq}_{n}\label{eq:noise3}
\end{equation}
and each $\epsilon_{n}^{pq}$ is a nonnegative parameter that reflects the maximum error expected in an element of the $n^{\rm th}$ shadow.  In this formulation we observe that the $N$-representability conditions serve not only to accelerate the convergence of the shadow tomography but also to mitigate errors in the shadows that potentially arise from the noise on a quantum device. The use of $N$-representability conditions for error mitigation in shadow tomography extends and generalizes earlier work in which $N$-representability conditions are applied post-measurement for error mitigation~\cite{Foley.2012, Rubin.2018, Smart.2019, Smart.2022w8u, Piskor.2023}. \\    


In the described program, the 2-positivity (DQG) conditions have a computational scaling of $r^4$ and $r^6$ in memory and floating-point operations, respectively~\cite{M2011}, where $r$ is the number of orbitals; therefore, the $N$-representability-enhanced tomography retains a polynomial scaling. Additionally, since the $N$-representability conditions are independent of a reference wave function, the described program applies to a wide range of strongly correlated systems. \\

\textit{Results.---}To implement shadow tomography with $N$-representability conditions, we solve the SDP in Eqs.~(\ref{eq:E}-\ref{eq:sdp-shadow}) by adding the shadow constraints to the variational 2-RDM method (v2RDM)~\cite{Mazziotti2004, Gidofalvi2005} in the Maple Quantum Chemistry Package~\cite{maple_2023, rdmchem_2023}.  We refer to the resulting algorithm as the shadow v2RDM (sv2RDM) method. The SDP is solved using the boundary-point algorithm in Ref.~\cite{M2011}. We generate the classical shadows from full configuration interaction (FCI) wave functions. The enhanced shadow tomography is applied to the dissociation of the nitrogen dimer and the ground-state energies of strongly correlated hydrogen chains~\cite{Suhai1994} with up to eight equally spaced atoms. The hydrogen atoms are represented in the minimal Slater-type-orbital (STO-3G) basis set~\cite{Hehre1969} while nitrogen is represented in the correlation-consistent polarized valence double-zeta (cc-pVDZ) basis set in a 10~electrons-in-8~orbitals [10,8] active space~\cite{Dunning1989}. \\

\begin{figure}[t!]
    \centering
    \includegraphics[width=\hsize]{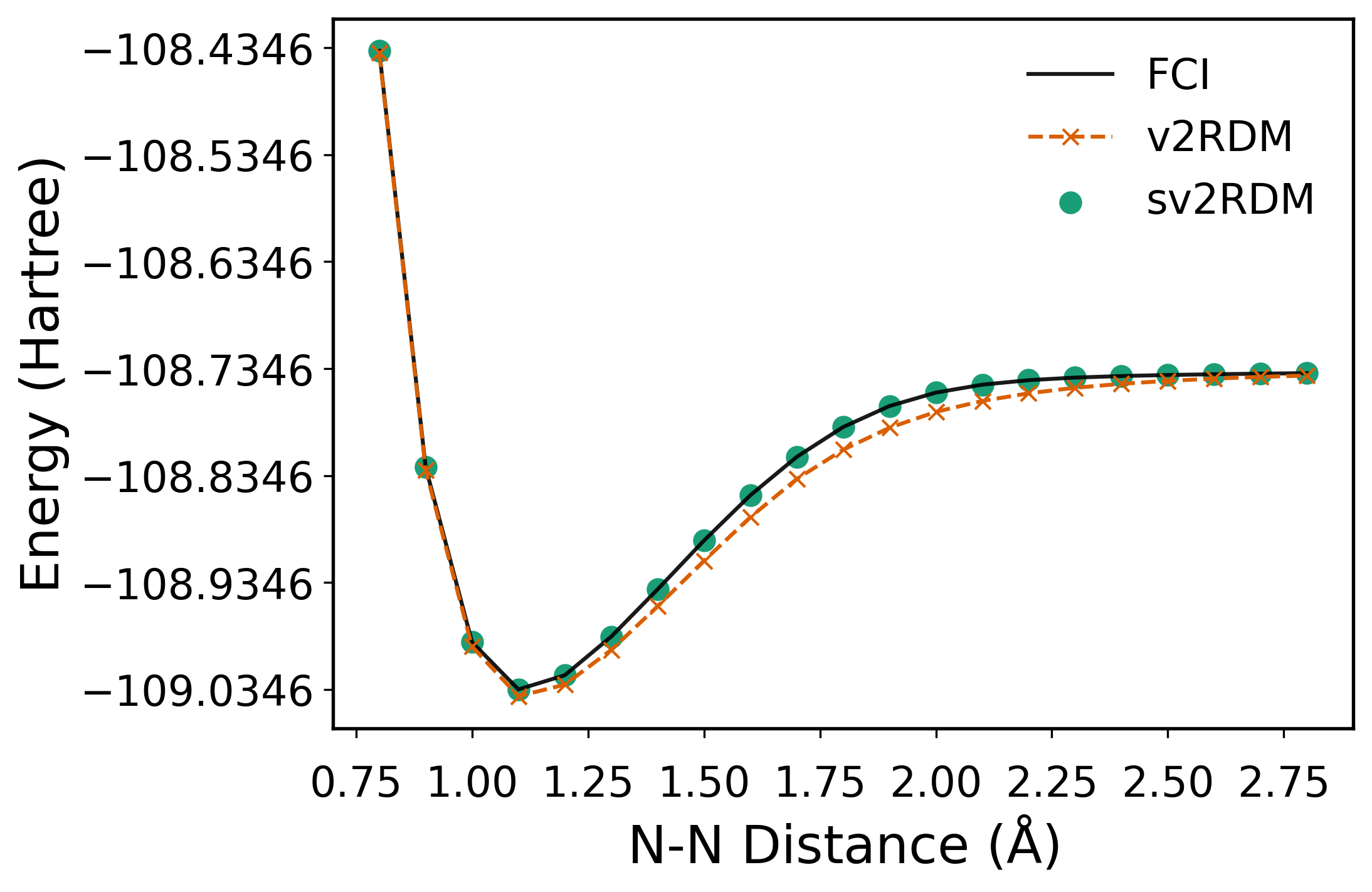}
    \caption{Potential energy surfaces of N$_2$ from FCI, v2RDM (DQG), and sv2RDM (DQG) are shown.  Total energies of sv2RDM with 18 shadows are in exact agreement with those calculated using FCI.}
    \label{fig:N2-PES}
\end{figure}

Total energies for N$_2$ from the sv2RDM method with 2-positivity (DQG) conditions converge quickly with the number of shadows to those from FCI as shown in Fig.~\ref{fig:N2-PES}. Note that for the FCI of N$_2$, when symmetry is neglected, the total number of configuration state functions is 1176~\cite{Cramer2013}. With 18 shadows, the Frobenius norm of the difference between the sv2RDM and FCI 2-RDMs is on the order of 10$^{-4}$. Hence, significantly fewer than $d^2$ shadows, where $d$ is the dimension of the 2-RDM describing the system, are needed for convergence to FCI energies and 2-RDMs, which is consistent with previous work on shadow tomography~\cite{Huang2020}. \\

\begin{figure}
    \centering
    \includegraphics[width=0.9\hsize]{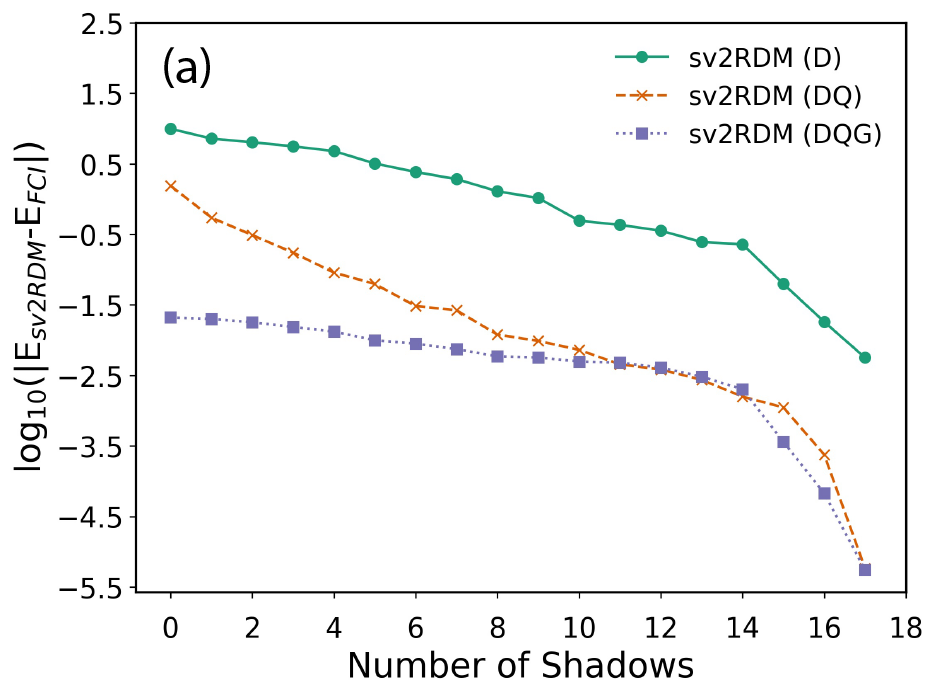}
    \includegraphics[width=0.9\hsize]{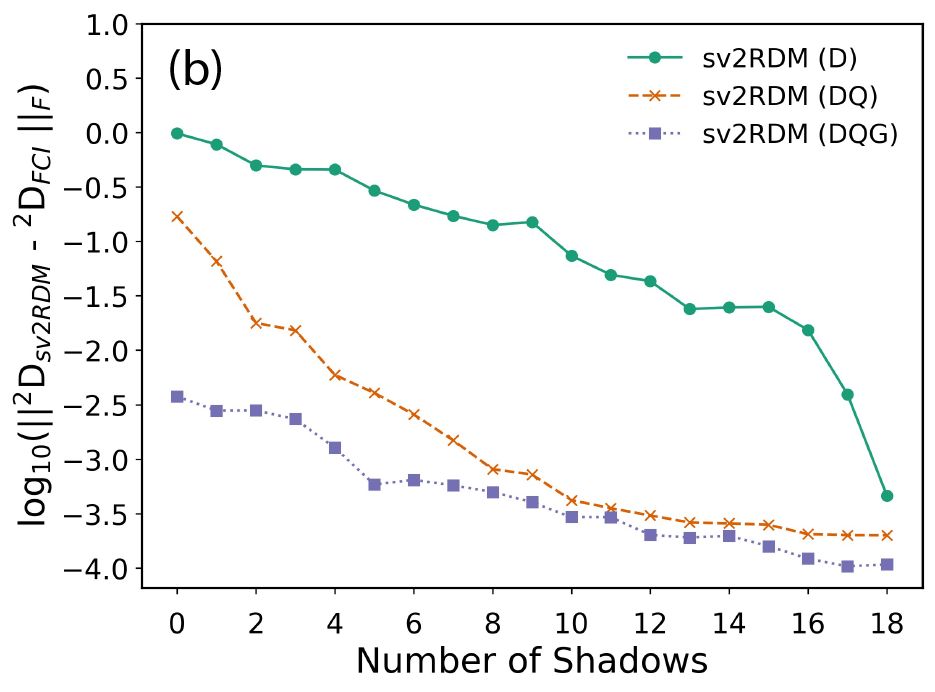}
    \caption{For N$_{2}$ at 1.75~\AA, the number of measurements required by the shadow tomography depends critically upon the number of $N$-representability conditions.  The energy and 2-RDM (normalized to one) errors with only 3 shadows and the DQ and the DQG conditions are more accurate than those with as many as 14 shadows and just the D condition.}
    \label{fig:N2-errors}
\end{figure}

For N$_{2}$ at 1.75~\AA, Fig.~\ref{fig:N2-errors} demonstrates that the number of measurements required by the shadow tomography depends critically upon the number of $N$-representability conditions.  The energy and 2-RDM errors with only 3 shadows and either the DQ or the DQG conditions are more accurate than those with as many as 14 shadows and just the D condition.  Moreover, the addition of the D condition is still better than shadow tomography without any positivity conditions.  For nearly all numbers of shadows in the range 0-17, the energy errors from DQG conditions are approximately two orders of magnitude better than those from just the D condition. Moreover, the 2-RDM errors from the DQG conditions are approximately three orders of magnitude better than those from the D condition. \\

While the D curve in both panels of Fig.~\ref{fig:N2-errors} decreases gradually in most of its domain, the DQ curve shows a steeper convergence pattern with respect to the number of shadows measured. This behavior indicates a nontrivial synergistic relationship between the classical shadows and the 2-RDM $N$-representability conditions that enhances the accuracy of the sampling. Overall, the total energy significantly relies on the positivity conditions, but as the energies and 2-RDMs of DQ approach those of DQG, the convergence slows and becomes more reliant on the shadow behavior. Therefore, considering $N$-representability conditions is crucial to achieving accurate electronic structure results and reliable convergence with the fewest number of shadows. \\

\begin{figure}
    \centering
    \includegraphics[width=0.95\hsize]{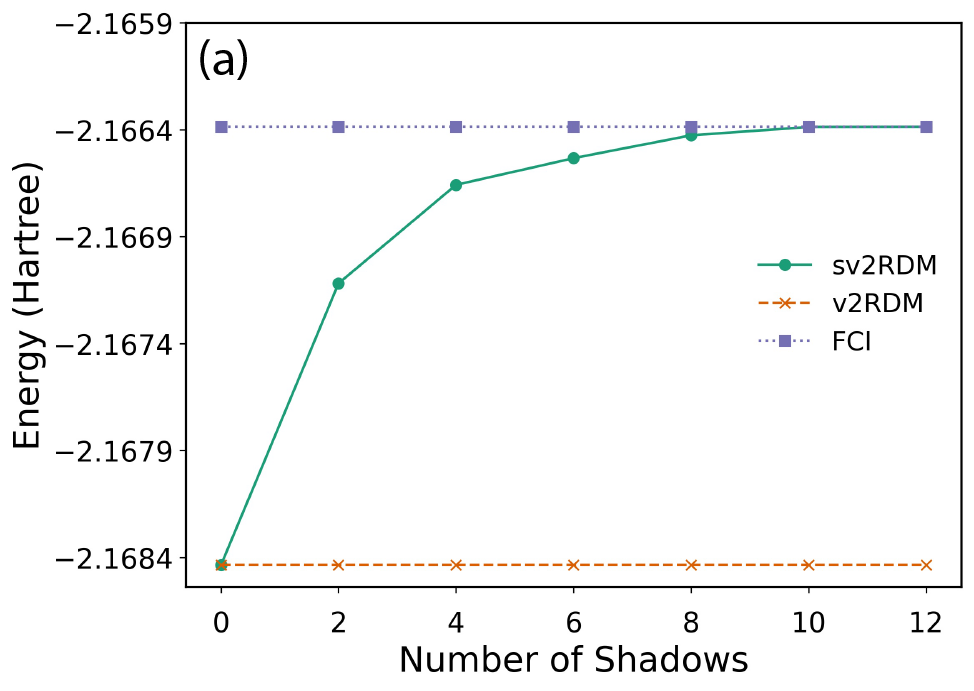}
    \includegraphics[width=0.95\hsize]{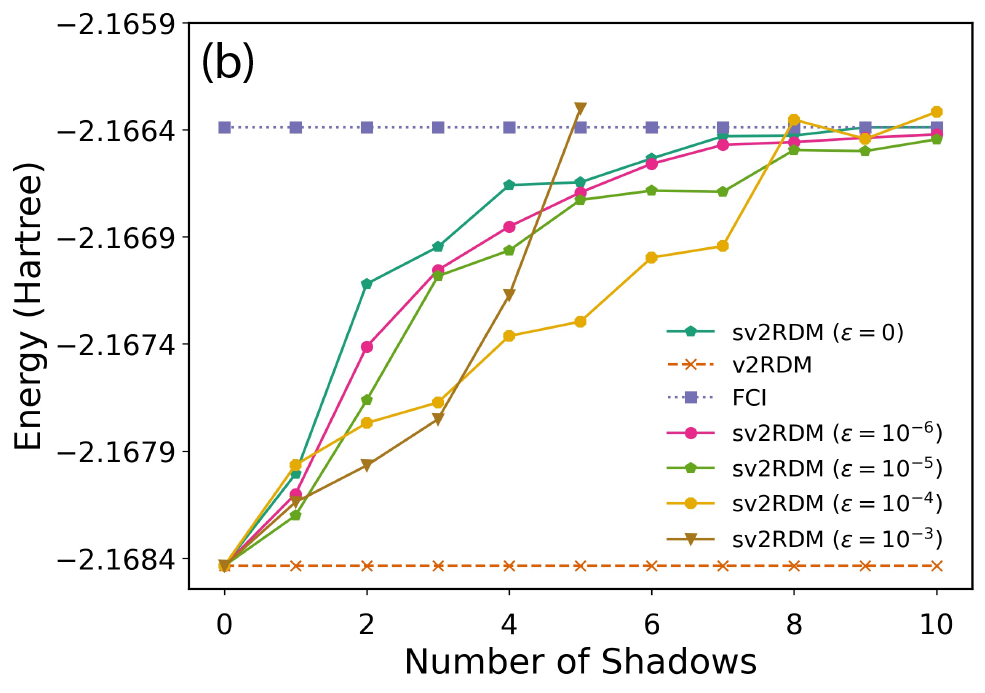}
    \caption{Total ground-state energy of H$_4$ with equally spaced H atoms is shown as a function of the number of shadows, (a) without and (b) with the presence of Gaussian noise.  The sv2RDM energy converges \textit{exactly} to the FCI energy within 10 shadows in the noiseless environment and \textit{approximately} in the noisy environments.}
    \label{fig:enegy-vs-shadows-H-chains-noise}
\end{figure}

The total energy as a function of the number of shadows for each hydrogen chain studied shows a logarithmic convergence pattern (Fig.~\ref{fig:enegy-vs-shadows-H-chains-noise}), with the Q and G conditions significantly improving the accuracy of the calculated values compared to the D condition alone (Table~\ref{H4_FCIerror}). With just 9 shadows, the sv2RDM total energy converges to the FCI energy of the H$_4$ chain in Fig.~\ref{fig:enegy-vs-shadows-H-chains-noise} (a). Additional numerical results on the hydrogen chains are available in Fig.~S1 and Tables~S1-S4 in the Supplemental Material (SM)~\footnote{The Supplemental Material (SM) presents additional data for longer hydrogen chains.} \\

Finally, we consider the performance of sv2RDM in the noisy setting where the measured 2-RDM contains errors introduced by a Gaussian noise matrix. The shadow constraints are replaced by the inequalities in Eqs.~(\ref{eq:noise1}-~\ref{eq:noise3}) that allow for errors $\epsilon_{n}^{pq}$. We choose $\epsilon_{n}^{pq} = \sigma$, where $\sigma$ is the standard deviation in the Gaussian noise model. Figure~\ref{fig:enegy-vs-shadows-H-chains-noise} (b) shows the sv2RDM convergence outcomes for various noise levels. Though the algorithmic performance is the best in low-noise environments, the example with even $\epsilon \leq$ 10$^{-4}$ shows the energy approaching that of the FCI. This result indicates that the $N$-representability conditions can help mitigate the errors introduced by noise and thereby retain chemical accuracy at a reduced measurement cost even in the presence of noise, which has potentially important applications on NISQ devices. \\

\begin{table}
\caption{\label{H4_FCIerror}For H$_4$ with 1.00 \r{A} H-H bonds the Frobenius norm of the 2-RDM (normalized to one) relative to FCI is presented as a function of the number of shadows and the number of $N$-representability conditions. The DQG conditions improve upon D by 1-3 orders of magnitude for all shadows.}
\begin{ruledtabular}
\begin{tabular}{cccc}
 Shadow number & D & DQ & DQG \\ 
  \hline
    1 & 0.21997267 & 0.02815293 & 0.00282400  \\ 
    3 & 0.21604040 & 0.00313894 & 0.00041823  \\ 
    5 & 0.01120927 & 0.00113098 & 0.00020163 \\ 
    7 & 0.00075773 & 0.00066025 & 0.00011213  \\ 
    9 & 0.00074105 & 0.00062567 & 0.00003492 \\ 
    11 & 0.00074105 & 0.00062914 & 0.00001140  \\ 
\end{tabular}
\end{ruledtabular}
\end{table}



\textit{Discussion and conclusions---} Shadow tomography has recently emerged as an important technique for characterizing quantum states, and yet despite its promise, it has potential limitations in its scalability for treating large quantum states and its sensitivity for dealing with quantum noise. We have demonstrated a method based on the theory of classical shadows that utilizes $N$-representability conditions to achieve the accuracy of target observables from the 2-RDM with fewer measurements than possible without these constraints. Our enhanced shadow tomography addresses the issue of exponential scaling of full quantum tomography by combining classical shadows for a succinct description of quantum states with the 2-RDM $N$-representability conditions, which allow for even fewer measurements with reliable convergence. The $N$-representability conditions are shown to decrease significantly the number of measurements needed to converge the 2-RDM for strongly correlated many-electron systems in both ideal and noisy settings, promising future advancements in many-body quantum tomography from reduced information. While the numerical calculations demonstrate exact agreement with electronic FCI wave functions, the presented theory is general and applicable to performing tomography of any exact or approximate many-body wave function. Future work will explore the ability of the algorithm to deal with different types of noise and its application to the quantum simulation of many-body problems on current quantum platforms.  \\




\begin{acknowledgments}

D.A.M. gratefully acknowledges the Department of Energy, Office of Basic Energy Sciences, Grant DE-SC0019215, the U.S. National Science Foundation Grant CHE-2155082 and DMR-2037783, and the NSF QuBBE Quantum Leap Challenge Institute (NSF OMA-2121044). I.A. gratefully acknowledges the NSF Graduate Research Fellowship Program under Grant No. 2140001. 

\end{acknowledgments}

\bibliography{references}

\begin{thebibliography}{74}%
\makeatletter
\providecommand \@ifxundefined [1]{%
 \@ifx{#1\undefined}
}%
\providecommand \@ifnum [1]{%
 \ifnum #1\expandafter \@firstoftwo
 \else \expandafter \@secondoftwo
 \fi
}%
\providecommand \@ifx [1]{%
 \ifx #1\expandafter \@firstoftwo
 \else \expandafter \@secondoftwo
 \fi
}%
\providecommand \natexlab [1]{#1}%
\providecommand \enquote  [1]{``#1''}%
\providecommand \bibnamefont  [1]{#1}%
\providecommand \bibfnamefont [1]{#1}%
\providecommand \citenamefont [1]{#1}%
\providecommand \href@noop [0]{\@secondoftwo}%
\providecommand \href [0]{\begingroup \@sanitize@url \@href}%
\providecommand \@href[1]{\@@startlink{#1}\@@href}%
\providecommand \@@href[1]{\endgroup#1\@@endlink}%
\providecommand \@sanitize@url [0]{\catcode `\\12\catcode `\$12\catcode
  `\&12\catcode `\#12\catcode `\^12\catcode `\_12\catcode `\%12\relax}%
\providecommand \@@startlink[1]{}%
\providecommand \@@endlink[0]{}%
\providecommand \url  [0]{\begingroup\@sanitize@url \@url }%
\providecommand \@url [1]{\endgroup\@href {#1}{\urlprefix }}%
\providecommand \urlprefix  [0]{URL }%
\providecommand \Eprint [0]{\href }%
\providecommand \doibase [0]{https://doi.org/}%
\providecommand \selectlanguage [0]{\@gobble}%
\providecommand \bibinfo  [0]{\@secondoftwo}%
\providecommand \bibfield  [0]{\@secondoftwo}%
\providecommand \translation [1]{[#1]}%
\providecommand \BibitemOpen [0]{}%
\providecommand \bibitemStop [0]{}%
\providecommand \bibitemNoStop [0]{.\EOS\space}%
\providecommand \EOS [0]{\spacefactor3000\relax}%
\providecommand \BibitemShut  [1]{\csname bibitem#1\endcsname}%
\let\auto@bib@innerbib\@empty
\bibitem [{\citenamefont {Aaronson}(2020)}]{Aaronson2020}%
  \BibitemOpen
  \bibfield  {author} {\bibinfo {author} {\bibfnamefont {S.}~\bibnamefont
  {Aaronson}},\ }\bibfield  {title} {\bibinfo {title} {Shadow tomography of
  quantum states},\ }\href {https://doi.org/10.1137/18m120275x} {\bibfield
  {journal} {\bibinfo  {journal} {SIAM Journal on Computing}\ }\textbf
  {\bibinfo {volume} {49}},\ \bibinfo {pages} {STOC18} (\bibinfo {year}
  {2020})}\BibitemShut {NoStop}%
\bibitem [{\citenamefont {Huang}\ \emph {et~al.}(2020)\citenamefont {Huang},
  \citenamefont {Kueng},\ and\ \citenamefont {Preskill}}]{Huang2020}%
  \BibitemOpen
  \bibfield  {author} {\bibinfo {author} {\bibfnamefont {H.-Y.}\ \bibnamefont
  {Huang}}, \bibinfo {author} {\bibfnamefont {R.}~\bibnamefont {Kueng}},\ and\
  \bibinfo {author} {\bibfnamefont {J.}~\bibnamefont {Preskill}},\ }\bibfield
  {title} {\bibinfo {title} {Predicting many properties of a quantum system
  from very few measurements},\ }\href
  {https://doi.org/10.1038/s41567-020-0932-7} {\bibfield  {journal} {\bibinfo
  {journal} {Nat. Phys}\ }\textbf {\bibinfo {volume} {16}},\ \bibinfo {pages}
  {1050} (\bibinfo {year} {2020})}\BibitemShut {NoStop}%
\bibitem [{\citenamefont {Huang}(2022)}]{Huang2022}%
  \BibitemOpen
  \bibfield  {author} {\bibinfo {author} {\bibfnamefont {H.-Y.}\ \bibnamefont
  {Huang}},\ }\bibfield  {title} {\bibinfo {title} {Learning quantum states
  from their classical shadows},\ }\href
  {https://doi.org/10.1038/s42254-021-00411-5} {\bibfield  {journal} {\bibinfo
  {journal} {Nat. Rev. Phys.}\ }\textbf {\bibinfo {volume} {4}},\ \bibinfo
  {pages} {81–81} (\bibinfo {year} {2022})}\BibitemShut {NoStop}%
\bibitem [{\citenamefont {Hu}\ \emph {et~al.}(2023)\citenamefont {Hu},
  \citenamefont {Choi},\ and\ \citenamefont {You}}]{Hu2023}%
  \BibitemOpen
  \bibfield  {author} {\bibinfo {author} {\bibfnamefont {H.-Y.}\ \bibnamefont
  {Hu}}, \bibinfo {author} {\bibfnamefont {S.}~\bibnamefont {Choi}},\ and\
  \bibinfo {author} {\bibfnamefont {Y.-Z.}\ \bibnamefont {You}},\ }\bibfield
  {title} {\bibinfo {title} {Classical shadow tomography with locally scrambled
  quantum dynamics},\ }\href {https://doi.org/10.1103/PhysRevResearch.5.023027}
  {\bibfield  {journal} {\bibinfo  {journal} {Phys. Rev. Res.}\ }\textbf
  {\bibinfo {volume} {5}},\ \bibinfo {pages} {023027} (\bibinfo {year}
  {2023})}\BibitemShut {NoStop}%
\bibitem [{\citenamefont {McGinley}\ \emph {et~al.}(2022)\citenamefont
  {McGinley}, \citenamefont {Leontica}, \citenamefont {Garratt}, \citenamefont
  {Jovanovic},\ and\ \citenamefont {Simon}}]{McGinley2022}%
  \BibitemOpen
  \bibfield  {author} {\bibinfo {author} {\bibfnamefont {M.}~\bibnamefont
  {McGinley}}, \bibinfo {author} {\bibfnamefont {S.}~\bibnamefont {Leontica}},
  \bibinfo {author} {\bibfnamefont {S.~J.}\ \bibnamefont {Garratt}}, \bibinfo
  {author} {\bibfnamefont {J.}~\bibnamefont {Jovanovic}},\ and\ \bibinfo
  {author} {\bibfnamefont {S.~H.}\ \bibnamefont {Simon}},\ }\bibfield  {title}
  {\bibinfo {title} {Quantifying information scrambling via classical shadow
  tomography on programmable quantum simulators},\ }\href
  {https://doi.org/10.1103/PhysRevA.106.012441} {\bibfield  {journal} {\bibinfo
   {journal} {Phys. Rev. A}\ }\textbf {\bibinfo {volume} {106}},\ \bibinfo
  {pages} {012441} (\bibinfo {year} {2022})}\BibitemShut {NoStop}%
\bibitem [{\citenamefont {O'Gorman}(2022)}]{OGorman2022}%
  \BibitemOpen
  \bibfield  {author} {\bibinfo {author} {\bibfnamefont {B.}~\bibnamefont
  {O'Gorman}},\ }\bibfield  {title} {\bibinfo {title} {{Fermionic tomography
  and learning}},\ }\bibfield  {journal} {\bibinfo  {journal} {arXiv}\ }\href
  {https://doi.org/10.48550/arxiv.2207.14787} {10.48550/arxiv.2207.14787}
  (\bibinfo {year} {2022})\BibitemShut {NoStop}%
\bibitem [{\citenamefont {Low}(2022)}]{Low2022}%
  \BibitemOpen
  \bibfield  {author} {\bibinfo {author} {\bibfnamefont {G.~H.}\ \bibnamefont
  {Low}},\ }\bibfield  {title} {\bibinfo {title} {{Classical Shadows of
  Fermions with Particle Number Symmetry}},\ }\bibfield  {journal} {\bibinfo
  {journal} {arXiv}\ }\href {https://doi.org/10.48550/arxiv.2208.08964}
  {10.48550/arxiv.2208.08964} (\bibinfo {year} {2022})\BibitemShut {NoStop}%
\bibitem [{\citenamefont {Zhao}\ \emph {et~al.}(2021)\citenamefont {Zhao},
  \citenamefont {Rubin},\ and\ \citenamefont {Miyake}}]{Zhao2021}%
  \BibitemOpen
  \bibfield  {author} {\bibinfo {author} {\bibfnamefont {A.}~\bibnamefont
  {Zhao}}, \bibinfo {author} {\bibfnamefont {N.~C.}\ \bibnamefont {Rubin}},\
  and\ \bibinfo {author} {\bibfnamefont {A.}~\bibnamefont {Miyake}},\
  }\bibfield  {title} {\bibinfo {title} {Fermionic partial tomography via
  classical shadows},\ }\href {https://doi.org/10.1103/physrevlett.127.110504}
  {\bibfield  {journal} {\bibinfo  {journal} {Phys. Rev. Lett.}\ }\textbf
  {\bibinfo {volume} {127}},\ \bibinfo {pages} {110504} (\bibinfo {year}
  {2021})}\BibitemShut {NoStop}%
\bibitem [{\citenamefont {Peng}\ \emph {et~al.}(2023)\citenamefont {Peng},
  \citenamefont {Zhang},\ and\ \citenamefont {Chan}}]{Peng2023}%
  \BibitemOpen
  \bibfield  {author} {\bibinfo {author} {\bibfnamefont {L.}~\bibnamefont
  {Peng}}, \bibinfo {author} {\bibfnamefont {X.}~\bibnamefont {Zhang}},\ and\
  \bibinfo {author} {\bibfnamefont {G.~K.-L.}\ \bibnamefont {Chan}},\
  }\bibfield  {title} {\bibinfo {title} {{Fermionic reduced density low-rank
  matrix completion, noise filtering, and measurement reduction in quantum
  simulations}},\ }\bibfield  {journal} {\bibinfo  {journal} {arXiv}\ }\href
  {https://doi.org/10.48550/arxiv.2306.05640} {10.48550/arxiv.2306.05640}
  (\bibinfo {year} {2023})\BibitemShut {NoStop}%
\bibitem [{\citenamefont {Lanyon}\ \emph {et~al.}(2017)\citenamefont {Lanyon},
  \citenamefont {Maier}, \citenamefont {Holz\"{a}pfel}, \citenamefont
  {Baumgratz}, \citenamefont {Hempel}, \citenamefont {Jurcevic}, \citenamefont
  {Dhand}, \citenamefont {Buyskikh}, \citenamefont {Daley}, \citenamefont
  {Cramer}, \citenamefont {Plenio}, \citenamefont {Blatt},\ and\ \citenamefont
  {Roos}}]{Lanyon2017}%
  \BibitemOpen
  \bibfield  {author} {\bibinfo {author} {\bibfnamefont {B.~P.}\ \bibnamefont
  {Lanyon}}, \bibinfo {author} {\bibfnamefont {C.}~\bibnamefont {Maier}},
  \bibinfo {author} {\bibfnamefont {M.}~\bibnamefont {Holz\"{a}pfel}}, \bibinfo
  {author} {\bibfnamefont {T.}~\bibnamefont {Baumgratz}}, \bibinfo {author}
  {\bibfnamefont {C.}~\bibnamefont {Hempel}}, \bibinfo {author} {\bibfnamefont
  {P.}~\bibnamefont {Jurcevic}}, \bibinfo {author} {\bibfnamefont
  {I.}~\bibnamefont {Dhand}}, \bibinfo {author} {\bibfnamefont {A.~S.}\
  \bibnamefont {Buyskikh}}, \bibinfo {author} {\bibfnamefont {A.~J.}\
  \bibnamefont {Daley}}, \bibinfo {author} {\bibfnamefont {M.}~\bibnamefont
  {Cramer}}, \bibinfo {author} {\bibfnamefont {M.~B.}\ \bibnamefont {Plenio}},
  \bibinfo {author} {\bibfnamefont {R.}~\bibnamefont {Blatt}},\ and\ \bibinfo
  {author} {\bibfnamefont {C.~F.}\ \bibnamefont {Roos}},\ }\bibfield  {title}
  {\bibinfo {title} {Efficient tomography of a quantum many-body system},\
  }\href {https://doi.org/10.1038/nphys4244} {\bibfield  {journal} {\bibinfo
  {journal} {Nat. Phys}\ }\textbf {\bibinfo {volume} {13}},\ \bibinfo {pages}
  {1158–1162} (\bibinfo {year} {2017})}\BibitemShut {NoStop}%
\bibitem [{\citenamefont {Yen}\ \emph {et~al.}(2020)\citenamefont {Yen},
  \citenamefont {Verteletskyi},\ and\ \citenamefont {Izmaylov}}]{Yen2020}%
  \BibitemOpen
  \bibfield  {author} {\bibinfo {author} {\bibfnamefont {T.-C.}\ \bibnamefont
  {Yen}}, \bibinfo {author} {\bibfnamefont {V.}~\bibnamefont {Verteletskyi}},\
  and\ \bibinfo {author} {\bibfnamefont {A.~F.}\ \bibnamefont {Izmaylov}},\
  }\bibfield  {title} {\bibinfo {title} {Measuring all compatible operators in
  one series of single-qubit measurements using unitary transformations},\
  }\href {https://doi.org/10.1021/acs.jctc.0c00008} {\bibfield  {journal}
  {\bibinfo  {journal} {J Theor Comput Chem.}\ }\textbf {\bibinfo {volume}
  {16}},\ \bibinfo {pages} {2400–2409} (\bibinfo {year} {2020})}\BibitemShut
  {NoStop}%
\bibitem [{\citenamefont {Reagor}\ \emph {et~al.}(2018)\citenamefont {Reagor}
  \emph {et~al.}}]{Reagor2018}%
  \BibitemOpen
  \bibfield  {author} {\bibinfo {author} {\bibfnamefont {M.}~\bibnamefont
  {Reagor}} \emph {et~al.},\ }\bibfield  {title} {\bibinfo {title}
  {Demonstration of universal parametric entangling gates on a multi-qubit
  lattice},\ }\href {https://doi.org/10.1126/sciadv.aao3603} {\bibfield
  {journal} {\bibinfo  {journal} {Sci. Adv.}\ }\textbf {\bibinfo {volume}
  {4}},\ \bibinfo {pages} {eaao3603} (\bibinfo {year} {2018})}\BibitemShut
  {NoStop}%
\bibitem [{\citenamefont {Cramer}\ \emph {et~al.}(2010)\citenamefont {Cramer},
  \citenamefont {Plenio}, \citenamefont {Flammia}, \citenamefont {Somma},
  \citenamefont {Gross}, \citenamefont {Bartlett}, \citenamefont
  {Landon-Cardinal}, \citenamefont {Poulin},\ and\ \citenamefont
  {Liu}}]{Cramer2010}%
  \BibitemOpen
  \bibfield  {author} {\bibinfo {author} {\bibfnamefont {M.}~\bibnamefont
  {Cramer}}, \bibinfo {author} {\bibfnamefont {M.~B.}\ \bibnamefont {Plenio}},
  \bibinfo {author} {\bibfnamefont {S.~T.}\ \bibnamefont {Flammia}}, \bibinfo
  {author} {\bibfnamefont {R.}~\bibnamefont {Somma}}, \bibinfo {author}
  {\bibfnamefont {D.}~\bibnamefont {Gross}}, \bibinfo {author} {\bibfnamefont
  {S.~D.}\ \bibnamefont {Bartlett}}, \bibinfo {author} {\bibfnamefont
  {O.}~\bibnamefont {Landon-Cardinal}}, \bibinfo {author} {\bibfnamefont
  {D.}~\bibnamefont {Poulin}},\ and\ \bibinfo {author} {\bibfnamefont {Y.-K.}\
  \bibnamefont {Liu}},\ }\bibfield  {title} {\bibinfo {title} {Efficient
  quantum state tomography},\ }\href {https://doi.org/10.1038/ncomms1147}
  {\bibfield  {journal} {\bibinfo  {journal} {Nat. Commun}\ }\textbf {\bibinfo
  {volume} {1}},\ \bibinfo {pages} {149} (\bibinfo {year} {2010})}\BibitemShut
  {NoStop}%
\bibitem [{\citenamefont {Rambach}\ \emph {et~al.}(2021)\citenamefont
  {Rambach}, \citenamefont {Qaryan}, \citenamefont {Kewming}, \citenamefont
  {Ferrie}, \citenamefont {White},\ and\ \citenamefont {Romero}}]{Rambach2021}%
  \BibitemOpen
  \bibfield  {author} {\bibinfo {author} {\bibfnamefont {M.}~\bibnamefont
  {Rambach}}, \bibinfo {author} {\bibfnamefont {M.}~\bibnamefont {Qaryan}},
  \bibinfo {author} {\bibfnamefont {M.}~\bibnamefont {Kewming}}, \bibinfo
  {author} {\bibfnamefont {C.}~\bibnamefont {Ferrie}}, \bibinfo {author}
  {\bibfnamefont {A.~G.}\ \bibnamefont {White}},\ and\ \bibinfo {author}
  {\bibfnamefont {J.}~\bibnamefont {Romero}},\ }\bibfield  {title} {\bibinfo
  {title} {Robust and efficient high-dimensional quantum state tomography},\
  }\href {https://doi.org/10.1103/PhysRevLett.126.100402} {\bibfield  {journal}
  {\bibinfo  {journal} {Phys. Rev. Lett.}\ }\textbf {\bibinfo {volume} {126}},\
  \bibinfo {pages} {100402} (\bibinfo {year} {2021})}\BibitemShut {NoStop}%
\bibitem [{\citenamefont {O’Donnell}\ and\ \citenamefont
  {Wright}(2016)}]{ODonnell2016}%
  \BibitemOpen
  \bibfield  {author} {\bibinfo {author} {\bibfnamefont {R.}~\bibnamefont
  {O’Donnell}}\ and\ \bibinfo {author} {\bibfnamefont {J.}~\bibnamefont
  {Wright}},\ }\bibfield  {title} {\bibinfo {title} {Efficient quantum
  tomography},\ }in\ \href {https://doi.org/10.1145/2897518.2897544} {\emph
  {\bibinfo {booktitle} {Proceedings of the forty-eighth annual ACM symposium
  on Theory of Computing}}},\ \bibinfo {series and number} {STOC ’16}\
  (\bibinfo  {publisher} {ACM},\ \bibinfo {year} {2016})\BibitemShut {NoStop}%
\bibitem [{\citenamefont {Haah}\ \emph {et~al.}(2017)\citenamefont {Haah},
  \citenamefont {Harrow}, \citenamefont {Ji}, \citenamefont {Wu},\ and\
  \citenamefont {Yu}}]{Haah2017}%
  \BibitemOpen
  \bibfield  {author} {\bibinfo {author} {\bibfnamefont {J.}~\bibnamefont
  {Haah}}, \bibinfo {author} {\bibfnamefont {A.~W.}\ \bibnamefont {Harrow}},
  \bibinfo {author} {\bibfnamefont {Z.}~\bibnamefont {Ji}}, \bibinfo {author}
  {\bibfnamefont {X.}~\bibnamefont {Wu}},\ and\ \bibinfo {author}
  {\bibfnamefont {N.}~\bibnamefont {Yu}},\ }\bibfield  {title} {\bibinfo
  {title} {Sample-optimal tomography of quantum states},\ }\href
  {https://doi.org/10.1109/TIT.2017.2719044} {\bibfield  {journal} {\bibinfo
  {journal} {IEEE Trans. Inf. Theory}\ }\textbf {\bibinfo {volume} {63}},\
  \bibinfo {pages} {5628} (\bibinfo {year} {2017})}\BibitemShut {NoStop}%
\bibitem [{\citenamefont {Gupta}\ \emph {et~al.}(2021)\citenamefont {Gupta},
  \citenamefont {Xia}, \citenamefont {Levine},\ and\ \citenamefont
  {Kais}}]{Gupta2021}%
  \BibitemOpen
  \bibfield  {author} {\bibinfo {author} {\bibfnamefont {R.}~\bibnamefont
  {Gupta}}, \bibinfo {author} {\bibfnamefont {R.}~\bibnamefont {Xia}}, \bibinfo
  {author} {\bibfnamefont {R.~D.}\ \bibnamefont {Levine}},\ and\ \bibinfo
  {author} {\bibfnamefont {S.}~\bibnamefont {Kais}},\ }\bibfield  {title}
  {\bibinfo {title} {Maximal entropy approach for quantum state tomography},\
  }\href {https://doi.org/10.1103/PRXQuantum.2.010318} {\bibfield  {journal}
  {\bibinfo  {journal} {PRX Quantum}\ }\textbf {\bibinfo {volume} {2}},\
  \bibinfo {pages} {010318} (\bibinfo {year} {2021})}\BibitemShut {NoStop}%
\bibitem [{\citenamefont {Gupta}\ \emph {et~al.}(2022)\citenamefont {Gupta},
  \citenamefont {Sajjan}, \citenamefont {Levine},\ and\ \citenamefont
  {Kais}}]{Gupta2022}%
  \BibitemOpen
  \bibfield  {author} {\bibinfo {author} {\bibfnamefont {R.}~\bibnamefont
  {Gupta}}, \bibinfo {author} {\bibfnamefont {M.}~\bibnamefont {Sajjan}},
  \bibinfo {author} {\bibfnamefont {R.~D.}\ \bibnamefont {Levine}},\ and\
  \bibinfo {author} {\bibfnamefont {S.}~\bibnamefont {Kais}},\ }\bibfield
  {title} {\bibinfo {title} {Variational approach to quantum state tomography
  based on maximal entropy formalism},\ }\href
  {https://doi.org/10.1039/d2cp04493e} {\bibfield  {journal} {\bibinfo
  {journal} {Phys. Chem. Chem. Phys.}\ }\textbf {\bibinfo {volume} {24}},\
  \bibinfo {pages} {28870–28877} (\bibinfo {year} {2022})}\BibitemShut
  {NoStop}%
\bibitem [{\citenamefont {Gard}\ \emph {et~al.}(2020)\citenamefont {Gard},
  \citenamefont {Zhu}, \citenamefont {Barron}, \citenamefont {Mayhall},
  \citenamefont {Economou},\ and\ \citenamefont {Barnes}}]{Gard2020}%
  \BibitemOpen
  \bibfield  {author} {\bibinfo {author} {\bibfnamefont {B.~T.}\ \bibnamefont
  {Gard}}, \bibinfo {author} {\bibfnamefont {L.}~\bibnamefont {Zhu}}, \bibinfo
  {author} {\bibfnamefont {G.~S.}\ \bibnamefont {Barron}}, \bibinfo {author}
  {\bibfnamefont {N.~J.}\ \bibnamefont {Mayhall}}, \bibinfo {author}
  {\bibfnamefont {S.~E.}\ \bibnamefont {Economou}},\ and\ \bibinfo {author}
  {\bibfnamefont {E.}~\bibnamefont {Barnes}},\ }\bibfield  {title} {\bibinfo
  {title} {Efficient symmetry-preserving state preparation circuits for the
  variational quantum eigensolver algorithm},\ }\href
  {https://doi.org/10.1038/s41534-019-0240-1} {\bibfield  {journal} {\bibinfo
  {journal} {Npj Quantum Inf.}\ }\textbf {\bibinfo {volume} {6}},\ \bibinfo
  {pages} {1} (\bibinfo {year} {2020})}\BibitemShut {NoStop}%
\bibitem [{\citenamefont {Tang}\ \emph {et~al.}(2021)\citenamefont {Tang},
  \citenamefont {Shkolnikov}, \citenamefont {Barron}, \citenamefont {Grimsley},
  \citenamefont {Mayhall}, \citenamefont {Barnes},\ and\ \citenamefont
  {Economou}}]{Tang2021}%
  \BibitemOpen
  \bibfield  {author} {\bibinfo {author} {\bibfnamefont {H.~L.}\ \bibnamefont
  {Tang}}, \bibinfo {author} {\bibfnamefont {V.}~\bibnamefont {Shkolnikov}},
  \bibinfo {author} {\bibfnamefont {G.~S.}\ \bibnamefont {Barron}}, \bibinfo
  {author} {\bibfnamefont {H.~R.}\ \bibnamefont {Grimsley}}, \bibinfo {author}
  {\bibfnamefont {N.~J.}\ \bibnamefont {Mayhall}}, \bibinfo {author}
  {\bibfnamefont {E.}~\bibnamefont {Barnes}},\ and\ \bibinfo {author}
  {\bibfnamefont {S.~E.}\ \bibnamefont {Economou}},\ }\bibfield  {title}
  {\bibinfo {title} {Qubit-adapt-vqe: An adaptive algorithm for constructing
  hardware-efficient ans\"atze on a quantum processor},\ }\href
  {https://doi.org/10.1103/PRXQuantum.2.020310} {\bibfield  {journal} {\bibinfo
   {journal} {PRX Quantum}\ }\textbf {\bibinfo {volume} {2}},\ \bibinfo {pages}
  {020310} (\bibinfo {year} {2021})}\BibitemShut {NoStop}%
\bibitem [{\citenamefont {Corcoles}\ \emph {et~al.}(2020)\citenamefont
  {Corcoles}, \citenamefont {Kandala}, \citenamefont {Javadi-Abhari},
  \citenamefont {McClure}, \citenamefont {Cross}, \citenamefont {Temme},
  \citenamefont {Nation}, \citenamefont {Steffen},\ and\ \citenamefont
  {Gambetta}}]{Corcoles.2020}%
  \BibitemOpen
  \bibfield  {author} {\bibinfo {author} {\bibfnamefont {A.~D.}\ \bibnamefont
  {Corcoles}}, \bibinfo {author} {\bibfnamefont {A.}~\bibnamefont {Kandala}},
  \bibinfo {author} {\bibfnamefont {A.}~\bibnamefont {Javadi-Abhari}}, \bibinfo
  {author} {\bibfnamefont {D.~T.}\ \bibnamefont {McClure}}, \bibinfo {author}
  {\bibfnamefont {A.~W.}\ \bibnamefont {Cross}}, \bibinfo {author}
  {\bibfnamefont {K.}~\bibnamefont {Temme}}, \bibinfo {author} {\bibfnamefont
  {P.~D.}\ \bibnamefont {Nation}}, \bibinfo {author} {\bibfnamefont
  {M.}~\bibnamefont {Steffen}},\ and\ \bibinfo {author} {\bibfnamefont {J.~M.}\
  \bibnamefont {Gambetta}},\ }\bibfield  {title} {\bibinfo {title} {{Challenges
  and Opportunities of Near-Term Quantum Computing Systems}},\ }\href
  {https://doi.org/10.1109/jproc.2019.2954005} {\bibfield  {journal} {\bibinfo
  {journal} {Proceedings of the IEEE}\ }\textbf {\bibinfo {volume} {108}},\
  \bibinfo {pages} {1338} (\bibinfo {year} {2020})},\ \Eprint
  {https://arxiv.org/abs/1910.02894} {1910.02894} \BibitemShut {NoStop}%
\bibitem [{\citenamefont {Jiang}\ \emph {et~al.}(2020)\citenamefont {Jiang},
  \citenamefont {Wang}, \citenamefont {Qian}, \citenamefont {Chen},
  \citenamefont {Chen}, \citenamefont {Lu}, \citenamefont {Xia}, \citenamefont
  {Song}, \citenamefont {Zhu},\ and\ \citenamefont {Ma}}]{Jiang.2020}%
  \BibitemOpen
  \bibfield  {author} {\bibinfo {author} {\bibfnamefont {X.}~\bibnamefont
  {Jiang}}, \bibinfo {author} {\bibfnamefont {K.}~\bibnamefont {Wang}},
  \bibinfo {author} {\bibfnamefont {K.}~\bibnamefont {Qian}}, \bibinfo {author}
  {\bibfnamefont {Z.}~\bibnamefont {Chen}}, \bibinfo {author} {\bibfnamefont
  {Z.}~\bibnamefont {Chen}}, \bibinfo {author} {\bibfnamefont {L.}~\bibnamefont
  {Lu}}, \bibinfo {author} {\bibfnamefont {L.}~\bibnamefont {Xia}}, \bibinfo
  {author} {\bibfnamefont {F.}~\bibnamefont {Song}}, \bibinfo {author}
  {\bibfnamefont {S.}~\bibnamefont {Zhu}},\ and\ \bibinfo {author}
  {\bibfnamefont {X.}~\bibnamefont {Ma}},\ }\bibfield  {title} {\bibinfo
  {title} {{Towards the standardization of quantum state verification using
  optimal strategies}},\ }\href {https://doi.org/10.1038/s41534-020-00317-7}
  {\bibfield  {journal} {\bibinfo  {journal} {{Npj Quantum Inf.}}\ }\textbf
  {\bibinfo {volume} {6}},\ \bibinfo {pages} {90} (\bibinfo {year} {2020})},\
  \Eprint {https://arxiv.org/abs/2002.00640} {2002.00640} \BibitemShut
  {NoStop}%
\bibitem [{\citenamefont {McArdle}\ \emph {et~al.}(2020)\citenamefont
  {McArdle}, \citenamefont {Endo}, \citenamefont {Aspuru-Guzik}, \citenamefont
  {Benjamin},\ and\ \citenamefont {Yuan}}]{mcardle_2020}%
  \BibitemOpen
  \bibfield  {author} {\bibinfo {author} {\bibfnamefont {S.}~\bibnamefont
  {McArdle}}, \bibinfo {author} {\bibfnamefont {S.}~\bibnamefont {Endo}},
  \bibinfo {author} {\bibfnamefont {A.}~\bibnamefont {Aspuru-Guzik}}, \bibinfo
  {author} {\bibfnamefont {S.~C.}\ \bibnamefont {Benjamin}},\ and\ \bibinfo
  {author} {\bibfnamefont {X.}~\bibnamefont {Yuan}},\ }\bibfield  {title}
  {\bibinfo {title} {{Quantum computational chemistry}},\ }\href
  {https://doi.org/10.1103/RevModPhys.92.015003} {\bibfield  {journal}
  {\bibinfo  {journal} {Rev. Mod. Phys.}\ }\textbf {\bibinfo {volume} {92}},\
  \bibinfo {pages} {015003} (\bibinfo {year} {2020})}\BibitemShut {NoStop}%
\bibitem [{\citenamefont {Head-Marsden}\ \emph {et~al.}(2021)\citenamefont
  {Head-Marsden}, \citenamefont {Flick}, \citenamefont {Ciccarino},\ and\
  \citenamefont {Narang}}]{Head-Marsden2020}%
  \BibitemOpen
  \bibfield  {author} {\bibinfo {author} {\bibfnamefont {K.}~\bibnamefont
  {Head-Marsden}}, \bibinfo {author} {\bibfnamefont {J.}~\bibnamefont {Flick}},
  \bibinfo {author} {\bibfnamefont {C.~J.}\ \bibnamefont {Ciccarino}},\ and\
  \bibinfo {author} {\bibfnamefont {P.}~\bibnamefont {Narang}},\ }\bibfield
  {title} {\bibinfo {title} {{Quantum Information and Algorithms for Correlated
  Quantum Matter}},\ }\href {https://doi.org/10.1021/acs.chemrev.0c00620}
  {\bibfield  {journal} {\bibinfo  {journal} {Chem. Rev.}\ }\textbf {\bibinfo
  {volume} {121}},\ \bibinfo {pages} {5} (\bibinfo {year} {2021})}\BibitemShut
  {NoStop}%
\bibitem [{\citenamefont {Smart}\ and\ \citenamefont
  {Mazziotti}(2021)}]{Smart2021_2}%
  \BibitemOpen
  \bibfield  {author} {\bibinfo {author} {\bibfnamefont {S.~E.}\ \bibnamefont
  {Smart}}\ and\ \bibinfo {author} {\bibfnamefont {D.~A.}\ \bibnamefont
  {Mazziotti}},\ }\bibfield  {title} {\bibinfo {title} {Quantum solver of
  contracted eigenvalue equations for scalable molecular simulations on quantum
  computing devices},\ }\href {https://doi.org/10.1103/PhysRevLett.126.070504}
  {\bibfield  {journal} {\bibinfo  {journal} {Phys. Rev. Lett.}\ }\textbf
  {\bibinfo {volume} {126}},\ \bibinfo {pages} {070504} (\bibinfo {year}
  {2021})}\BibitemShut {NoStop}%
\bibitem [{\citenamefont {Benavides-Riveros}\ \emph {et~al.}(2022)\citenamefont
  {Benavides-Riveros}, \citenamefont {Chen}, \citenamefont {Schilling},
  \citenamefont {Mantilla},\ and\ \citenamefont
  {Pittalis}}]{Benavides-Riveros.2022}%
  \BibitemOpen
  \bibfield  {author} {\bibinfo {author} {\bibfnamefont {C.~L.}\ \bibnamefont
  {Benavides-Riveros}}, \bibinfo {author} {\bibfnamefont {L.}~\bibnamefont
  {Chen}}, \bibinfo {author} {\bibfnamefont {C.}~\bibnamefont {Schilling}},
  \bibinfo {author} {\bibfnamefont {S.}~\bibnamefont {Mantilla}},\ and\
  \bibinfo {author} {\bibfnamefont {S.}~\bibnamefont {Pittalis}},\ }\bibfield
  {title} {\bibinfo {title} {{Excitations of Quantum Many-Body Systems via
  Purified Ensembles: A Unitary-Coupled-Cluster-Based Approach}},\ }\href
  {https://doi.org/10.1103/physrevlett.129.066401} {\bibfield  {journal}
  {\bibinfo  {journal} {Phys. Rev. Lett.}\ }\textbf {\bibinfo {volume} {129}},\
  \bibinfo {pages} {066401} (\bibinfo {year} {2022})},\ \Eprint
  {https://arxiv.org/abs/2201.10974} {2201.10974} \BibitemShut {NoStop}%
\bibitem [{\citenamefont {Mazziotti}(2007)}]{M2007}%
  \BibitemOpen
  \bibfield  {author} {\bibinfo {author} {\bibfnamefont {D.~A.}\ \bibnamefont
  {Mazziotti}},\ }\href@noop {} {\emph {\bibinfo {title}
  {Reduced-Density-Matrix Mechanics: With Application to Many-Electron Atoms
  and Molecule}}},\ Vol.\ \bibinfo {volume} {134}\ (\bibinfo  {publisher} {Adv.
  Chem. Phys.},\ \bibinfo {address} {{W}iley, {N}ew {Y}ork},\ \bibinfo {year}
  {2007})\BibitemShut {NoStop}%
\bibitem [{\citenamefont {Coleman}\ and\ \citenamefont
  {Yukalov}(2000)}]{Coleman2000}%
  \BibitemOpen
  \bibfield  {author} {\bibinfo {author} {\bibfnamefont {A.}~\bibnamefont
  {Coleman}}\ and\ \bibinfo {author} {\bibfnamefont {V.}~\bibnamefont
  {Yukalov}},\ }\href {https://books.google.com/books?id=l0hG98ylbh4C} {\emph
  {\bibinfo {title} {Reduced Density Matrices: Coulson’s Challenge}}},\
  Lecture Notes in Chemistry\ (\bibinfo  {publisher} {Springer Berlin
  Heidelberg},\ \bibinfo {year} {2000})\BibitemShut {NoStop}%
\bibitem [{\citenamefont {Coleman}(1963)}]{Coleman1963}%
  \BibitemOpen
  \bibfield  {author} {\bibinfo {author} {\bibfnamefont {A.~J.}\ \bibnamefont
  {Coleman}},\ }\bibfield  {title} {\bibinfo {title} {{Structure of Fermion
  Density Matrices}},\ }\href {https://doi.org/10.1103/RevModPhys.35.668}
  {\bibfield  {journal} {\bibinfo  {journal} {Rev. Mod. Phys.}\ }\textbf
  {\bibinfo {volume} {35}},\ \bibinfo {pages} {668} (\bibinfo {year}
  {1963})}\BibitemShut {NoStop}%
\bibitem [{\citenamefont {Garrod}\ and\ \citenamefont
  {Percus}(1964)}]{Garrod.1964}%
  \BibitemOpen
  \bibfield  {author} {\bibinfo {author} {\bibfnamefont {C.}~\bibnamefont
  {Garrod}}\ and\ \bibinfo {author} {\bibfnamefont {J.~K.}\ \bibnamefont
  {Percus}},\ }\bibfield  {title} {\bibinfo {title} {{Reduction of the
  $N$‐Particle Variational Problem}},\ }\href
  {https://doi.org/10.1063/1.1704098} {\bibfield  {journal} {\bibinfo
  {journal} {J. Math. Phys.}\ }\textbf {\bibinfo {volume} {5}},\ \bibinfo
  {pages} {1756} (\bibinfo {year} {1964})}\BibitemShut {NoStop}%
\bibitem [{\citenamefont {Kummer}(1967)}]{Kummer1967}%
  \BibitemOpen
  \bibfield  {author} {\bibinfo {author} {\bibfnamefont {H.}~\bibnamefont
  {Kummer}},\ }\bibfield  {title} {\bibinfo {title} {{$N$}-representability
  problem for reduced density matrices},\ }\href
  {https://doi.org/10.1063/1.1705122} {\bibfield  {journal} {\bibinfo
  {journal} {J. Math. Phys.}\ }\textbf {\bibinfo {volume} {8}},\ \bibinfo
  {pages} {2063} (\bibinfo {year} {1967})}\BibitemShut {NoStop}%
\bibitem [{\citenamefont {Erdahl}(1978)}]{Erdahl1978}%
  \BibitemOpen
  \bibfield  {author} {\bibinfo {author} {\bibfnamefont {R.~M.}\ \bibnamefont
  {Erdahl}},\ }\bibfield  {title} {\bibinfo {title} {Representability},\ }\href
  {https://doi.org/10.1002/qua.560130603} {\bibfield  {journal} {\bibinfo
  {journal} {Int. J. Quantum Chem.}\ }\textbf {\bibinfo {volume} {13}},\
  \bibinfo {pages} {697} (\bibinfo {year} {1978})}\BibitemShut {NoStop}%
\bibitem [{\citenamefont {Mazziotti}\ and\ \citenamefont
  {Erdahl}(2001)}]{mazziotti2001}%
  \BibitemOpen
  \bibfield  {author} {\bibinfo {author} {\bibfnamefont {D.~A.}\ \bibnamefont
  {Mazziotti}}\ and\ \bibinfo {author} {\bibfnamefont {R.~M.}\ \bibnamefont
  {Erdahl}},\ }\bibfield  {title} {\bibinfo {title} {Uncertainty relations and
  reduced density matrices: Mapping many-body quantum mechanics onto four
  particles},\ }\href {https://doi.org/10.1103/PhysRevA.63.042113} {\bibfield
  {journal} {\bibinfo  {journal} {Phys. Rev. A}\ }\textbf {\bibinfo {volume}
  {63}},\ \bibinfo {pages} {042113} (\bibinfo {year} {2001})}\BibitemShut
  {NoStop}%
\bibitem [{\citenamefont {Mazziotti}(2006)}]{Mazziotti2006_2}%
  \BibitemOpen
  \bibfield  {author} {\bibinfo {author} {\bibfnamefont {D.~A.}\ \bibnamefont
  {Mazziotti}},\ }\bibfield  {title} {\bibinfo {title} {Variational
  reduced-density-matrix method using three-particle {$N$}-representability
  conditions with application to many-electron molecules},\ }\href
  {https://doi.org/10.1103/PhysRevA.74.032501} {\bibfield  {journal} {\bibinfo
  {journal} {Phys. Rev. A}\ }\textbf {\bibinfo {volume} {74}},\ \bibinfo
  {pages} {032501} (\bibinfo {year} {2006})}\BibitemShut {NoStop}%
\bibitem [{\citenamefont {Mazziotti}(2012)}]{Mazziotti2012}%
  \BibitemOpen
  \bibfield  {author} {\bibinfo {author} {\bibfnamefont {D.~A.}\ \bibnamefont
  {Mazziotti}},\ }\bibfield  {title} {\bibinfo {title} {Structure of fermionic
  density matrices: Complete $n$-representability conditions},\ }\href
  {https://doi.org/10.1103/PhysRevLett.108.263002} {\bibfield  {journal}
  {\bibinfo  {journal} {Phys. Rev. Lett.}\ }\textbf {\bibinfo {volume} {108}},\
  \bibinfo {pages} {263002} (\bibinfo {year} {2012})}\BibitemShut {NoStop}%
\bibitem [{\citenamefont {Mazziotti}(2023)}]{Mazziotti2023}%
  \BibitemOpen
  \bibfield  {author} {\bibinfo {author} {\bibfnamefont {D.~A.}\ \bibnamefont
  {Mazziotti}},\ }\bibfield  {title} {\bibinfo {title} {Quantum many-body
  theory from a solution of the {$N$}-representability problem},\ }\href
  {https://doi.org/10.1103/PhysRevLett.130.153001} {\bibfield  {journal}
  {\bibinfo  {journal} {Phys. Rev. Lett.}\ }\textbf {\bibinfo {volume} {130}},\
  \bibinfo {pages} {153001} (\bibinfo {year} {2023})}\BibitemShut {NoStop}%
\bibitem [{\citenamefont {Vandenberghe}\ and\ \citenamefont
  {Boyd}(1996)}]{VB1996}%
  \BibitemOpen
  \bibfield  {author} {\bibinfo {author} {\bibfnamefont {L.}~\bibnamefont
  {Vandenberghe}}\ and\ \bibinfo {author} {\bibfnamefont {S.}~\bibnamefont
  {Boyd}},\ }\bibfield  {title} {\bibinfo {title} {Semidefinite programming},\
  }\href {https://doi.org/10.1137/1038003} {\bibfield  {journal} {\bibinfo
  {journal} {SIAM Review}\ }\textbf {\bibinfo {volume} {38}},\ \bibinfo {pages}
  {49} (\bibinfo {year} {1996})}\BibitemShut {NoStop}%
\bibitem [{\citenamefont {Mazziotti}(2004)}]{Mazziotti2004}%
  \BibitemOpen
  \bibfield  {author} {\bibinfo {author} {\bibfnamefont {D.~A.}\ \bibnamefont
  {Mazziotti}},\ }\bibfield  {title} {\bibinfo {title} {Realization of quantum
  chemistry without wave functions through first-order semidefinite
  programming},\ }\href {https://doi.org/10.1103/physrevlett.93.213001}
  {\bibfield  {journal} {\bibinfo  {journal} {Phys. Rev. Lett.}\ }\textbf
  {\bibinfo {volume} {93}},\ \bibinfo {pages} {213001} (\bibinfo {year}
  {2004})}\BibitemShut {NoStop}%
\bibitem [{\citenamefont {Mazziotti}(2011)}]{M2011}%
  \BibitemOpen
  \bibfield  {author} {\bibinfo {author} {\bibfnamefont {D.~A.}\ \bibnamefont
  {Mazziotti}},\ }\bibfield  {title} {\bibinfo {title} {Large-scale
  semidefinite programming for many-electron quantum mechanics},\ }\href
  {https://doi.org/10.1103/PhysRevLett.106.083001} {\bibfield  {journal}
  {\bibinfo  {journal} {Phys. Rev. Lett.}\ }\textbf {\bibinfo {volume} {106}},\
  \bibinfo {pages} {083001} (\bibinfo {year} {2011})}\BibitemShut {NoStop}%
\bibitem [{\citenamefont {Nakata}\ \emph {et~al.}(2001)\citenamefont {Nakata},
  \citenamefont {Nakatsuji}, \citenamefont {Ehara}, \citenamefont {Fukuda},
  \citenamefont {Nakata},\ and\ \citenamefont {Fujisawa}}]{Nakata.2001}%
  \BibitemOpen
  \bibfield  {author} {\bibinfo {author} {\bibfnamefont {M.}~\bibnamefont
  {Nakata}}, \bibinfo {author} {\bibfnamefont {H.}~\bibnamefont {Nakatsuji}},
  \bibinfo {author} {\bibfnamefont {M.}~\bibnamefont {Ehara}}, \bibinfo
  {author} {\bibfnamefont {M.}~\bibnamefont {Fukuda}}, \bibinfo {author}
  {\bibfnamefont {K.}~\bibnamefont {Nakata}},\ and\ \bibinfo {author}
  {\bibfnamefont {K.}~\bibnamefont {Fujisawa}},\ }\bibfield  {title} {\bibinfo
  {title} {{Variational calculations of fermion second-order reduced density
  matrices by semidefinite programming algorithm}},\ }\href
  {https://doi.org/10.1063/1.1360199} {\bibfield  {journal} {\bibinfo
  {journal} {J. Chem. Phys.}\ }\textbf {\bibinfo {volume} {114}},\ \bibinfo
  {pages} {8282} (\bibinfo {year} {2001})}\BibitemShut {NoStop}%
\bibitem [{\citenamefont {Mazziotti}(2002)}]{Mazziotti.2002kyg}%
  \BibitemOpen
  \bibfield  {author} {\bibinfo {author} {\bibfnamefont {D.~A.}\ \bibnamefont
  {Mazziotti}},\ }\bibfield  {title} {\bibinfo {title} {{Variational
  minimization of atomic and molecular ground-state energies via the
  two-particle reduced density matrix}},\ }\href
  {https://doi.org/10.1103/physreva.65.062511} {\bibfield  {journal} {\bibinfo
  {journal} {Phys. Rev. A}\ }\textbf {\bibinfo {volume} {65}},\ \bibinfo
  {pages} {062511} (\bibinfo {year} {2002})}\BibitemShut {NoStop}%
\bibitem [{\citenamefont {Zhao}\ \emph {et~al.}(2004)\citenamefont {Zhao},
  \citenamefont {Braams}, \citenamefont {Fukuda}, \citenamefont {Overton},\
  and\ \citenamefont {Percus}}]{Zhao.2004}%
  \BibitemOpen
  \bibfield  {author} {\bibinfo {author} {\bibfnamefont {Z.}~\bibnamefont
  {Zhao}}, \bibinfo {author} {\bibfnamefont {B.~J.}\ \bibnamefont {Braams}},
  \bibinfo {author} {\bibfnamefont {M.}~\bibnamefont {Fukuda}}, \bibinfo
  {author} {\bibfnamefont {M.~L.}\ \bibnamefont {Overton}},\ and\ \bibinfo
  {author} {\bibfnamefont {J.~K.}\ \bibnamefont {Percus}},\ }\bibfield  {title}
  {\bibinfo {title} {{The reduced density matrix method for electronic
  structure calculations and the role of three-index representability
  conditions}},\ }\href {https://doi.org/10.1063/1.1636721} {\bibfield
  {journal} {\bibinfo  {journal} {J. Chem. Phys.}\ }\textbf {\bibinfo {volume}
  {120}},\ \bibinfo {pages} {2095} (\bibinfo {year} {2004})}\BibitemShut
  {NoStop}%
\bibitem [{\citenamefont {Cancès}\ \emph {et~al.}(2006)\citenamefont
  {Cancès}, \citenamefont {Stoltz},\ and\ \citenamefont
  {Lewin}}]{Cances.2006}%
  \BibitemOpen
  \bibfield  {author} {\bibinfo {author} {\bibfnamefont {E.}~\bibnamefont
  {Cancès}}, \bibinfo {author} {\bibfnamefont {G.}~\bibnamefont {Stoltz}},\
  and\ \bibinfo {author} {\bibfnamefont {M.}~\bibnamefont {Lewin}},\ }\bibfield
   {title} {\bibinfo {title} {{The electronic ground-state energy problem: A
  new reduced density matrix approach}},\ }\href
  {https://doi.org/10.1063/1.2222358} {\bibfield  {journal} {\bibinfo
  {journal} {J. Chem. Phys.}\ }\textbf {\bibinfo {volume} {125}},\ \bibinfo
  {pages} {064101} (\bibinfo {year} {2006})}\BibitemShut {NoStop}%
\bibitem [{\citenamefont {Gidofalvi}\ and\ \citenamefont
  {Mazziotti}(2008)}]{Gidofalvi.2008}%
  \BibitemOpen
  \bibfield  {author} {\bibinfo {author} {\bibfnamefont {G.}~\bibnamefont
  {Gidofalvi}}\ and\ \bibinfo {author} {\bibfnamefont {D.~A.}\ \bibnamefont
  {Mazziotti}},\ }\bibfield  {title} {\bibinfo {title} {{Active-space
  two-electron reduced-density-matrix method: Complete active-space
  calculations without diagonalization of the N-electron Hamiltonian}},\ }\href
  {https://doi.org/10.1063/1.2983652} {\bibfield  {journal} {\bibinfo
  {journal} {J. Chem. Phys.}\ }\textbf {\bibinfo {volume} {129}},\ \bibinfo
  {pages} {134108} (\bibinfo {year} {2008})}\BibitemShut {NoStop}%
\bibitem [{\citenamefont {Shenvi}\ and\ \citenamefont
  {Izmaylov}(2010)}]{Shenvi.2010}%
  \BibitemOpen
  \bibfield  {author} {\bibinfo {author} {\bibfnamefont {N.}~\bibnamefont
  {Shenvi}}\ and\ \bibinfo {author} {\bibfnamefont {A.~F.}\ \bibnamefont
  {Izmaylov}},\ }\bibfield  {title} {\bibinfo {title} {{Active-Space
  N-Representability Constraints for Variational Two-Particle Reduced Density
  Matrix Calculations}},\ }\href
  {https://doi.org/10.1103/physrevlett.105.213003} {\bibfield  {journal}
  {\bibinfo  {journal} {Phys. Rev. Lett.}\ }\textbf {\bibinfo {volume} {105}},\
  \bibinfo {pages} {213003} (\bibinfo {year} {2010})}\BibitemShut {NoStop}%
\bibitem [{\citenamefont {Verstichel}\ \emph {et~al.}(2011)\citenamefont
  {Verstichel}, \citenamefont {Aggelen}, \citenamefont {Poelmans},\ and\
  \citenamefont {Neck}}]{Verstichel.2011}%
  \BibitemOpen
  \bibfield  {author} {\bibinfo {author} {\bibfnamefont {B.}~\bibnamefont
  {Verstichel}}, \bibinfo {author} {\bibfnamefont {H.~v.}\ \bibnamefont
  {Aggelen}}, \bibinfo {author} {\bibfnamefont {W.}~\bibnamefont {Poelmans}},\
  and\ \bibinfo {author} {\bibfnamefont {D.~V.}\ \bibnamefont {Neck}},\
  }\bibfield  {title} {\bibinfo {title} {{Variational Two-Particle Density
  Matrix Calculation for the Hubbard Model Below Half Filling Using
  Spin-Adapted Lifting Conditions}},\ }\href
  {https://doi.org/10.1103/physrevlett.108.213001} {\bibfield  {journal}
  {\bibinfo  {journal} {Phys. Rev. Lett.}\ }\textbf {\bibinfo {volume} {108}},\
  \bibinfo {pages} {213001} (\bibinfo {year} {2011})}\BibitemShut {NoStop}%
\bibitem [{\citenamefont {Baumgratz}\ and\ \citenamefont
  {Plenio}(2012)}]{Baumgratz.2012}%
  \BibitemOpen
  \bibfield  {author} {\bibinfo {author} {\bibfnamefont {T.}~\bibnamefont
  {Baumgratz}}\ and\ \bibinfo {author} {\bibfnamefont {M.~B.}\ \bibnamefont
  {Plenio}},\ }\bibfield  {title} {\bibinfo {title} {{Lower bounds for ground
  states of condensed matter systems}},\ }\href
  {https://doi.org/10.1088/1367-2630/14/2/023027} {\bibfield  {journal}
  {\bibinfo  {journal} {New J. Phys.}\ }\textbf {\bibinfo {volume} {14}},\
  \bibinfo {pages} {023027} (\bibinfo {year} {2012})}\BibitemShut {NoStop}%
\bibitem [{\citenamefont {Mazziotti}(2016)}]{Mazziotti.2016co}%
  \BibitemOpen
  \bibfield  {author} {\bibinfo {author} {\bibfnamefont {D.~A.}\ \bibnamefont
  {Mazziotti}},\ }\bibfield  {title} {\bibinfo {title} {{Enhanced Constraints
  for Accurate Lower Bounds on Many-Electron Quantum Energies from Variational
  Two-Electron Reduced Density Matrix Theory}},\ }\href
  {https://doi.org/10.1103/physrevlett.117.153001} {\bibfield  {journal}
  {\bibinfo  {journal} {Phys. Rev. Lett.}\ }\textbf {\bibinfo {volume} {117}},\
  \bibinfo {pages} {153001} (\bibinfo {year} {2016})}\BibitemShut {NoStop}%
\bibitem [{\citenamefont {Alcoba}\ \emph {et~al.}(2018)\citenamefont {Alcoba},
  \citenamefont {Torre}, \citenamefont {Lain}, \citenamefont {Massaccesi},
  \citenamefont {Oña}, \citenamefont {Honoré}, \citenamefont {Poelmans},
  \citenamefont {Neck}, \citenamefont {Bultinck},\ and\ \citenamefont
  {Baerdemacker}}]{Alcoba.2018}%
  \BibitemOpen
  \bibfield  {author} {\bibinfo {author} {\bibfnamefont {D.~R.}\ \bibnamefont
  {Alcoba}}, \bibinfo {author} {\bibfnamefont {A.}~\bibnamefont {Torre}},
  \bibinfo {author} {\bibfnamefont {L.}~\bibnamefont {Lain}}, \bibinfo {author}
  {\bibfnamefont {G.~E.}\ \bibnamefont {Massaccesi}}, \bibinfo {author}
  {\bibfnamefont {O.~B.}\ \bibnamefont {Oña}}, \bibinfo {author}
  {\bibfnamefont {E.~M.}\ \bibnamefont {Honoré}}, \bibinfo {author}
  {\bibfnamefont {W.}~\bibnamefont {Poelmans}}, \bibinfo {author}
  {\bibfnamefont {D.~V.}\ \bibnamefont {Neck}}, \bibinfo {author}
  {\bibfnamefont {P.}~\bibnamefont {Bultinck}},\ and\ \bibinfo {author}
  {\bibfnamefont {S.~D.}\ \bibnamefont {Baerdemacker}},\ }\bibfield  {title}
  {\bibinfo {title} {{Direct variational determination of the two-electron
  reduced density matrix for doubly occupied-configuration-interaction wave
  functions: The influence of three-index N-representability conditions}},\
  }\href {https://doi.org/10.1063/1.5008811} {\bibfield  {journal} {\bibinfo
  {journal} {J. Chem. Phys.}\ }\textbf {\bibinfo {volume} {148}},\ \bibinfo
  {pages} {024105} (\bibinfo {year} {2018})}\BibitemShut {NoStop}%
\bibitem [{\citenamefont {Mazziotti}(2020)}]{Mazziotti.20206vx}%
  \BibitemOpen
  \bibfield  {author} {\bibinfo {author} {\bibfnamefont {D.~A.}\ \bibnamefont
  {Mazziotti}},\ }\bibfield  {title} {\bibinfo {title} {{Dual-cone variational
  calculation of the two-electron reduced density matrix}},\ }\href
  {https://doi.org/10.1103/physreva.102.052819} {\bibfield  {journal} {\bibinfo
   {journal} {Phys. Rev. A}\ }\textbf {\bibinfo {volume} {102}},\ \bibinfo
  {pages} {052819} (\bibinfo {year} {2020})}\BibitemShut {NoStop}%
\bibitem [{\citenamefont {Li}\ \emph {et~al.}(2021)\citenamefont {Li},
  \citenamefont {Liebenthal},\ and\ \citenamefont {DePrince}}]{Li.2021}%
  \BibitemOpen
  \bibfield  {author} {\bibinfo {author} {\bibfnamefont {R.~R.}\ \bibnamefont
  {Li}}, \bibinfo {author} {\bibfnamefont {M.~D.}\ \bibnamefont {Liebenthal}},\
  and\ \bibinfo {author} {\bibfnamefont {A.~E.}\ \bibnamefont {DePrince}},\
  }\bibfield  {title} {\bibinfo {title} {{Challenges for variational
  reduced-density-matrix theory with three-particle N-representability
  conditions}},\ }\href {https://doi.org/10.1063/5.0066404} {\bibfield
  {journal} {\bibinfo  {journal} {J. Chem. Phys.}\ }\textbf {\bibinfo {volume}
  {155}},\ \bibinfo {pages} {174110} (\bibinfo {year} {2021})}\BibitemShut
  {NoStop}%
\bibitem [{\citenamefont {Piris}(2021)}]{Piris.2021}%
  \BibitemOpen
  \bibfield  {author} {\bibinfo {author} {\bibfnamefont {M.}~\bibnamefont
  {Piris}},\ }\bibfield  {title} {\bibinfo {title} {{Global Natural Orbital
  Functional: Towards the Complete Description of the Electron Correlation}},\
  }\href {https://doi.org/10.1103/physrevlett.127.233001} {\bibfield  {journal}
  {\bibinfo  {journal} {Phys. Rev. Lett.}\ }\textbf {\bibinfo {volume} {127}},\
  \bibinfo {pages} {233001} (\bibinfo {year} {2021})},\ \Eprint
  {https://arxiv.org/abs/2112.02119} {2112.02119} \BibitemShut {NoStop}%
\bibitem [{\citenamefont {Knight}\ \emph {et~al.}(2022)\citenamefont {Knight},
  \citenamefont {Quiney},\ and\ \citenamefont {Martin}}]{Knight.2022}%
  \BibitemOpen
  \bibfield  {author} {\bibinfo {author} {\bibfnamefont {M.~J.}\ \bibnamefont
  {Knight}}, \bibinfo {author} {\bibfnamefont {H.~M.}\ \bibnamefont {Quiney}},\
  and\ \bibinfo {author} {\bibfnamefont {A.~M.}\ \bibnamefont {Martin}},\
  }\bibfield  {title} {\bibinfo {title} {{Reduced density matrix approach to
  ultracold few-fermion systems in one dimension}},\ }\href
  {https://doi.org/10.1088/1367-2630/ac643d} {\bibfield  {journal} {\bibinfo
  {journal} {New J. Phys.}\ }\textbf {\bibinfo {volume} {24}},\ \bibinfo
  {pages} {053004} (\bibinfo {year} {2022})},\ \Eprint
  {https://arxiv.org/abs/2106.09187} {2106.09187} \BibitemShut {NoStop}%
\bibitem [{\citenamefont {Schlimgen}\ \emph {et~al.}(2016)\citenamefont
  {Schlimgen}, \citenamefont {Heaps},\ and\ \citenamefont
  {Mazziotti}}]{Schlimgen.2016}%
  \BibitemOpen
  \bibfield  {author} {\bibinfo {author} {\bibfnamefont {A.~W.}\ \bibnamefont
  {Schlimgen}}, \bibinfo {author} {\bibfnamefont {C.~W.}\ \bibnamefont
  {Heaps}},\ and\ \bibinfo {author} {\bibfnamefont {D.~A.}\ \bibnamefont
  {Mazziotti}},\ }\bibfield  {title} {\bibinfo {title} {{Entangled Electrons
  Foil Synthesis of Elusive Low-Valent Vanadium Oxo Complex}},\ }\href
  {https://doi.org/10.1021/acs.jpclett.5b02547} {\bibfield  {journal} {\bibinfo
   {journal} {J. Phys. Chem. Lett.}\ }\textbf {\bibinfo {volume} {7}},\
  \bibinfo {pages} {627} (\bibinfo {year} {2016})}\BibitemShut {NoStop}%
\bibitem [{\citenamefont {Boyn}\ \emph {et~al.}(2020)\citenamefont {Boyn},
  \citenamefont {Xie}, \citenamefont {Anderson},\ and\ \citenamefont
  {Mazziotti}}]{Boyn.2020}%
  \BibitemOpen
  \bibfield  {author} {\bibinfo {author} {\bibfnamefont {J.-N.}\ \bibnamefont
  {Boyn}}, \bibinfo {author} {\bibfnamefont {J.}~\bibnamefont {Xie}}, \bibinfo
  {author} {\bibfnamefont {J.~S.}\ \bibnamefont {Anderson}},\ and\ \bibinfo
  {author} {\bibfnamefont {D.~A.}\ \bibnamefont {Mazziotti}},\ }\bibfield
  {title} {\bibinfo {title} {{Entangled Electrons Drive a Non-superexchange
  Mechanism in a Cobalt Quinoid Dimer Complex}},\ }\href
  {https://doi.org/10.1021/acs.jpclett.0c01248} {\bibfield  {journal} {\bibinfo
   {journal} {J. Phys. Chem. Lett.}\ }\textbf {\bibinfo {volume} {11}},\
  \bibinfo {pages} {4584} (\bibinfo {year} {2020})},\ \Eprint
  {https://arxiv.org/abs/2005.03637} {2005.03637} \BibitemShut {NoStop}%
\bibitem [{\citenamefont {Kawamura}\ \emph {et~al.}(2020)\citenamefont
  {Kawamura}, \citenamefont {Xie}, \citenamefont {Boyn}, \citenamefont {Jesse},
  \citenamefont {McNeece}, \citenamefont {Hill}, \citenamefont {Collins},
  \citenamefont {Valdez-Moreira}, \citenamefont {Filatov}, \citenamefont
  {Kurutz}, \citenamefont {Mazziotti},\ and\ \citenamefont
  {Anderson}}]{Kawamura.2020jms}%
  \BibitemOpen
  \bibfield  {author} {\bibinfo {author} {\bibfnamefont {A.}~\bibnamefont
  {Kawamura}}, \bibinfo {author} {\bibfnamefont {J.}~\bibnamefont {Xie}},
  \bibinfo {author} {\bibfnamefont {J.-N.}\ \bibnamefont {Boyn}}, \bibinfo
  {author} {\bibfnamefont {K.~A.}\ \bibnamefont {Jesse}}, \bibinfo {author}
  {\bibfnamefont {A.~J.}\ \bibnamefont {McNeece}}, \bibinfo {author}
  {\bibfnamefont {E.~A.}\ \bibnamefont {Hill}}, \bibinfo {author}
  {\bibfnamefont {K.~A.}\ \bibnamefont {Collins}}, \bibinfo {author}
  {\bibfnamefont {J.~A.}\ \bibnamefont {Valdez-Moreira}}, \bibinfo {author}
  {\bibfnamefont {A.~S.}\ \bibnamefont {Filatov}}, \bibinfo {author}
  {\bibfnamefont {J.~W.}\ \bibnamefont {Kurutz}}, \bibinfo {author}
  {\bibfnamefont {D.~A.}\ \bibnamefont {Mazziotti}},\ and\ \bibinfo {author}
  {\bibfnamefont {J.~S.}\ \bibnamefont {Anderson}},\ }\bibfield  {title}
  {\bibinfo {title} {{Reversible Switching of Organic Diradical Character via
  Iron-Based Spin-Crossover}},\ }\href {https://doi.org/10.1021/jacs.0c08307}
  {\bibfield  {journal} {\bibinfo  {journal} {J. Am. Chem. Soc.}\ }\textbf
  {\bibinfo {volume} {142}},\ \bibinfo {pages} {17670} (\bibinfo {year}
  {2020})}\BibitemShut {NoStop}%
\bibitem [{\citenamefont {Xie}\ \emph {et~al.}(2022)\citenamefont {Xie},
  \citenamefont {Ewing}, \citenamefont {Boyn}, \citenamefont {Filatov},
  \citenamefont {Cheng}, \citenamefont {Ma}, \citenamefont {Grocke},
  \citenamefont {Zhao}, \citenamefont {Itani}, \citenamefont {Sun},
  \citenamefont {Cho}, \citenamefont {Chen}, \citenamefont {Chapman},
  \citenamefont {Patel}, \citenamefont {Talapin}, \citenamefont {Park},
  \citenamefont {Mazziotti},\ and\ \citenamefont {Anderson}}]{Xie.20228s}%
  \BibitemOpen
  \bibfield  {author} {\bibinfo {author} {\bibfnamefont {J.}~\bibnamefont
  {Xie}}, \bibinfo {author} {\bibfnamefont {S.}~\bibnamefont {Ewing}}, \bibinfo
  {author} {\bibfnamefont {J.-N.}\ \bibnamefont {Boyn}}, \bibinfo {author}
  {\bibfnamefont {A.~S.}\ \bibnamefont {Filatov}}, \bibinfo {author}
  {\bibfnamefont {B.}~\bibnamefont {Cheng}}, \bibinfo {author} {\bibfnamefont
  {T.}~\bibnamefont {Ma}}, \bibinfo {author} {\bibfnamefont {G.~L.}\
  \bibnamefont {Grocke}}, \bibinfo {author} {\bibfnamefont {N.}~\bibnamefont
  {Zhao}}, \bibinfo {author} {\bibfnamefont {R.}~\bibnamefont {Itani}},
  \bibinfo {author} {\bibfnamefont {X.}~\bibnamefont {Sun}}, \bibinfo {author}
  {\bibfnamefont {H.}~\bibnamefont {Cho}}, \bibinfo {author} {\bibfnamefont
  {Z.}~\bibnamefont {Chen}}, \bibinfo {author} {\bibfnamefont {K.~W.}\
  \bibnamefont {Chapman}}, \bibinfo {author} {\bibfnamefont {S.~N.}\
  \bibnamefont {Patel}}, \bibinfo {author} {\bibfnamefont {D.~V.}\ \bibnamefont
  {Talapin}}, \bibinfo {author} {\bibfnamefont {J.}~\bibnamefont {Park}},
  \bibinfo {author} {\bibfnamefont {D.~A.}\ \bibnamefont {Mazziotti}},\ and\
  \bibinfo {author} {\bibfnamefont {J.~S.}\ \bibnamefont {Anderson}},\
  }\bibfield  {title} {\bibinfo {title} {{Intrinsic glassy-metallic transport
  in an amorphous coordination polymer}},\ }\href
  {https://doi.org/10.1038/s41586-022-05261-4} {\bibfield  {journal} {\bibinfo
  {journal} {Nature}\ }\textbf {\bibinfo {volume} {611}},\ \bibinfo {pages}
  {479} (\bibinfo {year} {2022})}\BibitemShut {NoStop}%
\bibitem [{\citenamefont {Schouten}\ \emph {et~al.}(2023)\citenamefont
  {Schouten}, \citenamefont {Klevens}, \citenamefont {Sager-Smith},
  \citenamefont {Xie}, \citenamefont {Anderson},\ and\ \citenamefont
  {Mazziotti}}]{Schouten.2023}%
  \BibitemOpen
  \bibfield  {author} {\bibinfo {author} {\bibfnamefont {A.~O.}\ \bibnamefont
  {Schouten}}, \bibinfo {author} {\bibfnamefont {J.~E.}\ \bibnamefont
  {Klevens}}, \bibinfo {author} {\bibfnamefont {L.~M.}\ \bibnamefont
  {Sager-Smith}}, \bibinfo {author} {\bibfnamefont {J.}~\bibnamefont {Xie}},
  \bibinfo {author} {\bibfnamefont {J.~S.}\ \bibnamefont {Anderson}},\ and\
  \bibinfo {author} {\bibfnamefont {D.~A.}\ \bibnamefont {Mazziotti}},\
  }\bibfield  {title} {\bibinfo {title} {{Potential for exciton condensation in
  a highly conductive amorphous polymer}},\ }\href
  {https://doi.org/10.1103/physrevmaterials.7.045001} {\bibfield  {journal}
  {\bibinfo  {journal} {Phys. Rev. Mat.}\ }\textbf {\bibinfo {volume} {7}},\
  \bibinfo {pages} {045001} (\bibinfo {year} {2023})}\BibitemShut {NoStop}%
\bibitem [{\citenamefont {Haar}(1933)}]{Haar1933}%
  \BibitemOpen
  \bibfield  {author} {\bibinfo {author} {\bibfnamefont {A.}~\bibnamefont
  {Haar}},\ }\bibfield  {title} {\bibinfo {title} {Der massbegriff in der
  theorie der kontinuierlichen gruppen},\ }\href
  {https://doi.org/10.2307/1968346} {\bibfield  {journal} {\bibinfo  {journal}
  {The Annals of Mathematics}\ }\textbf {\bibinfo {volume} {34}},\ \bibinfo
  {pages} {147} (\bibinfo {year} {1933})}\BibitemShut {NoStop}%
\bibitem [{\citenamefont {Candès}\ and\ \citenamefont
  {Recht}(2009)}]{Candes.2009}%
  \BibitemOpen
  \bibfield  {author} {\bibinfo {author} {\bibfnamefont {E.~J.}\ \bibnamefont
  {Candès}}\ and\ \bibinfo {author} {\bibfnamefont {B.}~\bibnamefont
  {Recht}},\ }\bibfield  {title} {\bibinfo {title} {{Exact Matrix Completion
  via Convex Optimization}},\ }\href
  {https://doi.org/10.1007/s10208-009-9045-5} {\bibfield  {journal} {\bibinfo
  {journal} {Found. Comput. Math.}\ }\textbf {\bibinfo {volume} {9}},\ \bibinfo
  {pages} {717} (\bibinfo {year} {2009})}\BibitemShut {NoStop}%
\bibitem [{\citenamefont {Cai}\ \emph {et~al.}(2010)\citenamefont {Cai},
  \citenamefont {Cands},\ and\ \citenamefont {Shen}}]{Cai.2010}%
  \BibitemOpen
  \bibfield  {author} {\bibinfo {author} {\bibfnamefont {J.-F.}\ \bibnamefont
  {Cai}}, \bibinfo {author} {\bibfnamefont {E.~J.}\ \bibnamefont {Cands}},\
  and\ \bibinfo {author} {\bibfnamefont {Z.}~\bibnamefont {Shen}},\ }\bibfield
  {title} {\bibinfo {title} {{A Singular Value Thresholding Algorithm for
  Matrix Completion}},\ }\href {https://doi.org/10.1137/080738970} {\bibfield
  {journal} {\bibinfo  {journal} {SIAM J. Optim.}\ }\textbf {\bibinfo {volume}
  {20}},\ \bibinfo {pages} {1956} (\bibinfo {year} {2010})}\BibitemShut
  {NoStop}%
\bibitem [{\citenamefont {Foley}\ and\ \citenamefont
  {Mazziotti}(2012)}]{Foley.2012}%
  \BibitemOpen
  \bibfield  {author} {\bibinfo {author} {\bibfnamefont {J.~J.}\ \bibnamefont
  {Foley}}\ and\ \bibinfo {author} {\bibfnamefont {D.~A.}\ \bibnamefont
  {Mazziotti}},\ }\bibfield  {title} {\bibinfo {title} {{Measurement-driven
  reconstruction of many-particle quantum processes by semidefinite programming
  with application to photosynthetic light harvesting}},\ }\href
  {https://doi.org/10.1103/physreva.86.012512} {\bibfield  {journal} {\bibinfo
  {journal} {Phys. Rev. A}\ }\textbf {\bibinfo {volume} {86}},\ \bibinfo
  {pages} {012512} (\bibinfo {year} {2012})}\BibitemShut {NoStop}%
\bibitem [{\citenamefont {Rubin}\ \emph {et~al.}(2018)\citenamefont {Rubin},
  \citenamefont {Babbush},\ and\ \citenamefont {McClean}}]{Rubin.2018}%
  \BibitemOpen
  \bibfield  {author} {\bibinfo {author} {\bibfnamefont {N.~C.}\ \bibnamefont
  {Rubin}}, \bibinfo {author} {\bibfnamefont {R.}~\bibnamefont {Babbush}},\
  and\ \bibinfo {author} {\bibfnamefont {J.}~\bibnamefont {McClean}},\
  }\bibfield  {title} {\bibinfo {title} {{Application of fermionic marginal
  constraints to hybrid quantum algorithms}},\ }\href
  {https://doi.org/10.1088/1367-2630/aab919} {\bibfield  {journal} {\bibinfo
  {journal} {New J. Phys.}\ }\textbf {\bibinfo {volume} {20}},\ \bibinfo
  {pages} {053020} (\bibinfo {year} {2018})}\BibitemShut {NoStop}%
\bibitem [{\citenamefont {Smart}\ and\ \citenamefont
  {Mazziotti}(2019)}]{Smart.2019}%
  \BibitemOpen
  \bibfield  {author} {\bibinfo {author} {\bibfnamefont {S.~E.}\ \bibnamefont
  {Smart}}\ and\ \bibinfo {author} {\bibfnamefont {D.~A.}\ \bibnamefont
  {Mazziotti}},\ }\bibfield  {title} {\bibinfo {title} {{Quantum-classical
  hybrid algorithm using an error-mitigating N-representability condition to
  compute the Mott metal-insulator transition}},\ }\href
  {https://doi.org/10.1103/physreva.100.022517} {\bibfield  {journal} {\bibinfo
   {journal} {Phys. Rev. A}\ }\textbf {\bibinfo {volume} {100}},\ \bibinfo
  {pages} {022517} (\bibinfo {year} {2019})},\ \Eprint
  {https://arxiv.org/abs/2004.07739} {2004.07739} \BibitemShut {NoStop}%
\bibitem [{\citenamefont {Smart}\ \emph {et~al.}(2022)\citenamefont {Smart},
  \citenamefont {Boyn},\ and\ \citenamefont {Mazziotti}}]{Smart.2022w8u}%
  \BibitemOpen
  \bibfield  {author} {\bibinfo {author} {\bibfnamefont {S.~E.}\ \bibnamefont
  {Smart}}, \bibinfo {author} {\bibfnamefont {J.-N.}\ \bibnamefont {Boyn}},\
  and\ \bibinfo {author} {\bibfnamefont {D.~A.}\ \bibnamefont {Mazziotti}},\
  }\bibfield  {title} {\bibinfo {title} {{Resolving correlated states of
  benzyne with an error-mitigated contracted quantum eigensolver}},\ }\href
  {https://doi.org/10.1103/physreva.105.022405} {\bibfield  {journal} {\bibinfo
   {journal} {Phys. Rev. A}\ }\textbf {\bibinfo {volume} {105}},\ \bibinfo
  {pages} {022405} (\bibinfo {year} {2022})},\ \Eprint
  {https://arxiv.org/abs/2103.06876} {2103.06876} \BibitemShut {NoStop}%
\bibitem [{\citenamefont {Piskor}\ \emph {et~al.}(2023)\citenamefont {Piskor},
  \citenamefont {Eich}, \citenamefont {Marthaler}, \citenamefont {Wilhelm},\
  and\ \citenamefont {Reiner}}]{Piskor.2023}%
  \BibitemOpen
  \bibfield  {author} {\bibinfo {author} {\bibfnamefont {T.}~\bibnamefont
  {Piskor}}, \bibinfo {author} {\bibfnamefont {F.~G.}\ \bibnamefont {Eich}},
  \bibinfo {author} {\bibfnamefont {M.}~\bibnamefont {Marthaler}}, \bibinfo
  {author} {\bibfnamefont {F.~K.}\ \bibnamefont {Wilhelm}},\ and\ \bibinfo
  {author} {\bibfnamefont {J.-M.}\ \bibnamefont {Reiner}},\ }\bibfield  {title}
  {\bibinfo {title} {{Post-processing noisy quantum computations utilizing
  $N$-representability constraints}},\ }\bibfield  {journal} {\bibinfo
  {journal} {arXiv}\ }\href {https://doi.org/10.48550/arxiv.2304.13401}
  {10.48550/arxiv.2304.13401} (\bibinfo {year} {2023}),\ \Eprint
  {https://arxiv.org/abs/2304.13401} {2304.13401} \BibitemShut {NoStop}%
\bibitem [{\citenamefont {Gidofalvi}\ and\ \citenamefont
  {Mazziotti}(2005)}]{Gidofalvi2005}%
  \BibitemOpen
  \bibfield  {author} {\bibinfo {author} {\bibfnamefont {G.}~\bibnamefont
  {Gidofalvi}}\ and\ \bibinfo {author} {\bibfnamefont {D.~A.}\ \bibnamefont
  {Mazziotti}},\ }\bibfield  {title} {\bibinfo {title} {Spin and symmetry
  adaptation of the variational two-electron reduced-density-matrix method},\
  }\href {https://doi.org/10.1103/PhysRevA.72.052505} {\bibfield  {journal}
  {\bibinfo  {journal} {Phys. Rev. A}\ }\textbf {\bibinfo {volume} {72}},\
  \bibinfo {pages} {052505} (\bibinfo {year} {2005})}\BibitemShut {NoStop}%
\bibitem [{map(2023)}]{maple_2023}%
  \BibitemOpen
  \href {https://www.maplesoft.com} {\emph {\bibinfo {title} {Maple}}}\
  (\bibinfo  {publisher} {Maplesoft, Waterloo},\ \bibinfo {year}
  {2023})\BibitemShut {NoStop}%
\bibitem [{rdm(2023)}]{rdmchem_2023}%
  \BibitemOpen
  \href {https://www.rdmchem.com} {\emph {\bibinfo {title} {RDMChem, Quantum
  Chemistry Package}}}\ (\bibinfo  {publisher} {Maplesoft, Waterloo},\ \bibinfo
  {year} {2023})\BibitemShut {NoStop}%
\bibitem [{\citenamefont {Suhai}(1994)}]{Suhai1994}%
  \BibitemOpen
  \bibfield  {author} {\bibinfo {author} {\bibfnamefont {S.}~\bibnamefont
  {Suhai}},\ }\bibfield  {title} {\bibinfo {title} {Electron correlation in
  extended systems: Fourth-order many-body perturbation theory and
  density-functional methods applied to an infinite chain of hydrogen atoms},\
  }\href {https://doi.org/10.1103/PhysRevB.50.14791} {\bibfield  {journal}
  {\bibinfo  {journal} {Phys. Rev. B}\ }\textbf {\bibinfo {volume} {50}},\
  \bibinfo {pages} {14791} (\bibinfo {year} {1994})}\BibitemShut {NoStop}%
\bibitem [{\citenamefont {Hehre}\ \emph {et~al.}(1969)\citenamefont {Hehre},
  \citenamefont {Stewart},\ and\ \citenamefont {Pople}}]{Hehre1969}%
  \BibitemOpen
  \bibfield  {author} {\bibinfo {author} {\bibfnamefont {W.~J.}\ \bibnamefont
  {Hehre}}, \bibinfo {author} {\bibfnamefont {R.~F.}\ \bibnamefont {Stewart}},\
  and\ \bibinfo {author} {\bibfnamefont {J.~A.}\ \bibnamefont {Pople}},\
  }\bibfield  {title} {\bibinfo {title} {Self-consistent molecular-orbital
  methods. i. use of gaussian expansions of slater-type atomic orbitals},\
  }\href {https://doi.org/10.1063/1.1672392} {\bibfield  {journal} {\bibinfo
  {journal} {Chem. Phys.}\ }\textbf {\bibinfo {volume} {51}},\ \bibinfo {pages}
  {2657–2664} (\bibinfo {year} {1969})}\BibitemShut {NoStop}%
\bibitem [{\citenamefont {Dunning}(1989)}]{Dunning1989}%
  \BibitemOpen
  \bibfield  {author} {\bibinfo {author} {\bibfnamefont {T.~H.}\ \bibnamefont
  {Dunning}},\ }\bibfield  {title} {\bibinfo {title} {Gaussian basis sets for
  use in correlated molecular calculations. i. the atoms boron through neon and
  hydrogen},\ }\href {https://doi.org/10.1063/1.456153} {\bibfield  {journal}
  {\bibinfo  {journal} {Chem. Phys.}\ }\textbf {\bibinfo {volume} {90}},\
  \bibinfo {pages} {1007–1023} (\bibinfo {year} {1989})}\BibitemShut
  {NoStop}%
\bibitem [{\citenamefont {Cramer}(2013)}]{Cramer2013}%
  \BibitemOpen
  \bibfield  {author} {\bibinfo {author} {\bibfnamefont {C.~J.}\ \bibnamefont
  {Cramer}},\ }\href@noop {} {\emph {\bibinfo {title} {Essentials of
  Computational Chemistry: Theories and Models}}}\ (\bibinfo  {publisher} {John
  Wiley \& Sons},\ \bibinfo {year} {2013})\BibitemShut {NoStop}%
\bibitem [{Note1()}]{Note1}%
  \BibitemOpen
  \bibinfo {note} {The Supplemental Material (SM) presents additional data for
  longer hydrogen chains.}\BibitemShut {Stop}%
\end{thebibliography}%

\end{document}